\documentclass{aa}

\usepackage{txfonts}
\usepackage[bookmarks=true,
            bookmarksopen=true,
            pdfstartview = FitV,
            pdfpagelayout = SinglePage,
            colorlinks = true,
            linkcolor=black,
            urlcolor = blue,
            pdfborder = {0 0 0}]{hyperref}

\usepackage{orcidlink}

\begin{document}

   \title{A possible two-fold scenario for the disc-corona of the luminous AGN 1H\,0419--577: a high-density disc or a warm corona}

   \author{D.\ Porquet \inst{1}\orcidlink{0000-0001-9731-0352}
         \and
         J.~N.\ Reeves\inst{2,3}\orcidlink{0000-0003-3221-6765}
            \and
          V.\ Braito\inst{3,2}\orcidlink{0000-0002-2629-4989}
         }
   \institute{Aix Marseille Univ, CNRS, CNES, LAM, Marseille, France 
              \email{delphine.porquet@lam.fr}
        \and  Department of Physics, Institute for Astrophysics and Computational
Sciences, The Catholic University of America, Washington, DC 20064, USA
          \and INAF, Osservatorio Astronomico di Brera, Via Bianchi 46 I-23807 Merate
 (LC), Italy
              }
   \date{Received , 2025; accepted , 2025}

 \titlerunning{A possible two-fold scenario for the disc-corona of the luminous highly-accreting AGN 1H\,0419--577}
 \authorrunning{Porquet et al.}

  \abstract
  {1H\,0419–577 is a highly-accreting, luminous broad-line type-I active galactic nucleus (AGN). The process(es) at work in its disc-corona system, especially the origin of the soft X-ray excess, is still highly debated based on {\sl XMM-Newton} observations: relativistic reflection from the illumination of the accretion disc by the hot corona versus Comptonisation of seed photons from the accretion disc by a warm corona, in addition to the hot one.}
   {This study aims to characterise its disc-corona system using, for the first time, simultaneous \textit{XMM-Newton} and \textit{NuSTAR} observations, performed in May and November 2018.}
   {We conducted high-resolution grating spectroscopy to identify potential soft X-ray absorption and emission features. To measure the hot corona temperatures from the spectral analysis above 3 keV, we also included data from a previous \textit{NuSTAR} observation from June 2015. We characterised the disc-corona system properties by analysing the broadband spectra and the spectral energy distribution (SED) from UV to hard X-rays.}
   {1H 0419-577 was observed in a \textit{bare-like} high-flux state at both epochs, with negligible neutral and ionised absorption along its line of sight at both Galactic and AGN rest-frames. However, several soft X-ray emission lines were detected, notably a broad and intense \ion{O}{vii} line indicating an accretion disc origin at only a few tens of gravitational radii. The broadband X-ray spectra revealed a prominent, absorption-free smooth soft X-ray excess, a weak Fe\,K$\alpha$ complex, and a lack of a Compton hump. Fitting data above 3 keV yielded apparent moderate hot corona temperatures of $\sim$20--30 keV for the 2018 and 2015 observations, depending on the model applied. The 2018 X-ray broadband spectra were well reproduced by either a relativistic reflection model with a high-density accretion disc ($\sim$10$^{18}$\,cm$^{-2}$), or a hybrid model combining warm and hot coronae with relativistic reflection. We performed the SED analysis for the latter scenario, which indicated that both the hot and warm coronae would have a small spatial extent.}
    {Both scenarios can successfully reproduce the two 2018 observations of 1H\,0419--577, but they imply very different physical conditions, for example, in terms of disc density, temperature and accretion power released in the hot corona  and the origin of the UV emission.}
   \maketitle
%

\section{Introduction}\label{sec:Introduction}

\begin{table*}[!t]
\caption{Observation log of the {\sl XMM-Newton} and {\sl NuSTAR} datasets for 1H\,0419-577 analysed in this work.}
\centering
\begin{tabular}{c@{\hspace{15pt}}c@{\hspace{15pt}}c@{\hspace{15pt}}c@{\hspace{15pt}}c@{\hspace{15pt}}c@{\hspace{15pt}}c@{\hspace{15pt}}c}
\hline \hline
\multicolumn{1}{c}{Mission} & \multicolumn{1}{c}{Obs.\,ID} & \multicolumn{1}{c}{Obs. date} & \multicolumn{1}{c}{Instr.} & \multicolumn{1}{c}{Exp.$^a$}  &  \multicolumn{1}{c}{Mean CR$^b$} \\
        &          &  \multicolumn{1}{c}{(yyyy-mm-dd)} & &   \multicolumn{1}{c}{(ks)} & \multicolumn{1}{c}{count\,s$^{-1}$} \\
\hline 
{\sl NuSTAR}     & 60402006002 & 2018-05-15  & FPMA & 64.2 & 0.29 \\
                 &             &             & FPMB & 64.0 & 0.28\\
{\sl XMM-Newton} & 0820360101  & 2018-05-16  & \multicolumn{1}{c}{pn} & 50.1 (35.1) & 8.7 \\
\hline
{\sl NuSTAR}     &  60402006004 & 2018-11-13 & FPMA & 48.3 &  0.34 \\
                 &              &            & FPMB & 48.0 & 0.32\\
{\sl XMM-Newton} & 0820360201 & 2018-11-13 & \multicolumn{1}{c}{pn} & 51.4 (34.1) & 9.8 \\
\hline
{\sl NuSTAR}     &  60101039002 & 2015-06-03  & FPMA & 169.5 &  0.37\\
                &               &             & FPMB & 169.1 &  0.36 \\
\hline
\end{tabular}
\label{tab:log}
\flushleft
\small{\textit{Notes.} $^a$ Total exposure and in parentheses the net pn exposure time after correction of background flares and deadtime (29\% for the small window mode). (b) Mean source count rate over 0.3–10 keV for {\sl XMM-Newton} and 3–79 keV for {\sl NuSTAR}.}
\end{table*}

\object{1H\,0419-577} ($z$=0.104) is a very bright radio-quiet broad-line active galactic nucleus (AGN) with a strong big blue bump \citep{Brissenden87,Turner99,Guainazzi98}. It has a full width at half maximum (FWHM) for H$\beta$ of 3241$\pm$366\,km\,s$^{-1}$ \citep{ChenY22}. The central supermassive black hole mass has been estimated to be log($M_{\sl BH}/M_{\odot}$)=8.123$^{+0.072}_{-0.087}$ from single-epoch spectroscopy of UV emission lines \citep{Tilton13}, with an Eddington accretion rate of about 0.1--0.4 \citep[e.g.,][]{Pounds04a,DiGesu14}, depending on its flux state. 
In X-rays, 1H\,0419--577 exhibits significant flux variability and spectral changes, with its soft X-ray flux varying by approximately a factor of ten over several years \citep{Guainazzi98,Pounds04a,DiGesu14}. Using multi-epoch short-length \textit{XMM-Newton} observations from 2002 and 2003, the variable soft X-ray excess strength has been attributed to variable absorption by low-ionisation matter close to the supermassive black hole \citep{Pounds04b}. This matter becomes more ionised and less opaque as the flux increases, while its covering factor ($\sim$50--60\%) remains almost constant \citep{DiGesu14}. Furthermore, the high angular resolution of \textit{Chandra} has revealed a $\sim$kpc extended X-ray outflow \citep{DiGesu17}. 
From its first \textit{NuSTAR} observation performed in 2015 ($\sim$160\,ks), a low-to-moderate hot corona temperature has been inferred, depending on the model applied ($kT_{\rm hot}$=15$\pm$1\,keV \citealt{TurnerJ18}; $kT_{\rm hot}$=30$^{+22}_{-7}$\,keV \citealt{JiangJ19a}).
Regarding the prominent soft X-ray excess observed for 1H\,0419-577, two alternative physical origins have been proposed: (i) a combination of light-bending from relativistic reflection and a thin warm absorber component \citep{Fabian05,JiangJ19b}, and (ii) a warm corona in addition to the standard hot corona \citep{DiGesu14,Petrucci18,Palit24,Ballantyne24}. However, these two scenarios have been tested solely on \textit{XMM-Newton} observations. Therefore, it is insightful to test them using simultaneous high signal-to-noise \textit{XMM-Newton} and \textit{NuSTAR} observations, and to infer the physical properties of the disc-corona system of 1H\,0419-577.

This work presents a detailed spectral analysis -- X-ray broadband and spectral energy distribution (SED) -- of the first two simultaneous \textit{XMM-Newton} and \textit{NuSTAR} observations of 1H\,0419-577, performed approximately six months apart in May and November 2018. The source was observed at both epochs in a high X-ray flux state. 
In Section~\ref{sec:obs}, we detail the data reduction and analysis methods for the dataset. Section~\ref{sec:RGS} presents the data analysis of the resolution grating spectrometer (RGS), showing that 1H\,0419-577 was observed in a bare-like state at both epochs with no noticeable rest-frame neutral or ionised absorber. The analysis of data above 3\,keV, including the long 2015 \textit{NuSTAR} observation, to probe the hot corona and relativistic reflection contributions, is reported in Section~\ref{sec:above3keV}.
The full X-ray broadband spectra are then investigated to test two alternative scenarios: the relativistic reflection solely (Section~\ref{sec:reflkerrd-0.3-79keV}) and the hybrid scenario including the presence of a warm corona in addition to the hot corona and relativistic reflection (Section~\ref{sec:warmcorona}). For the latter scenario, the SED is also used to constrain the physical parameters of the corresponding disc-corona system. Finally, the main results are summarised and discussed in Section~\ref{sec:discussion}.

\section{Observations, Data Reduction, and Analysis Method}\label{sec:obs}

\subsection{\textit{XMM-Newton} and \textit{NuSTAR} data reduction}

\textit{XMM-Newton} \citep{Jansen01} and \textit{NuSTAR} \citep{Harrison13} simultaneously observed 1H\,0419--577 for the first time in May and November 2018. For the data analysis above 3\,keV (see Section~\ref{sec:above3keV}), we also use the deep archival \textit{NuSTAR} observation performed in June 2015. Table~\ref{tab:log} summarises the main information about the datasets used in this work.

\subsubsection{\textit{XMM-Newton} Data Reduction}\label{sec:xmm}

The \textit{XMM-Newton} data were reprocessed using the Science Analysis System (\texttt{SAS}, version 21.0.0) with the most recent calibration as of September 3, 2024. Due to the high luminosity of 1H 0419-577, the European Photon Imaging Camera (EPIC) instruments, pn \citep{Struder01} and MOS \citep{Turner01}, were used in small window and partial window modes, respectively. This study focused solely on pn data, selecting event patterns 0–4 (single and double events) due to its superior sensitivity above about 6\,keV. A circular region with a 35$\arcsec$ radius, centered on 1H 0419-577, was used to extract pn spectra, avoiding the chip's edge. Background spectra were obtained from a rectangular area in the small window's lower section, containing minimal or no photons from 1H 0419-577. Table~\ref{tab:log} presents the total net exposure times, adjusted for deadtime (29\% for small window mode) and background flaring. 
The \texttt{SAS} tasks \texttt{rmfgen} and \texttt{arfgen} were employed to generate all redistribution matrix files (RMF) and ancillary response files (ARF). They were binned to oversample the instrumental resolution by at least a factor of four, with no impact on the fit results. For the arf calculation, we applied the recent option \texttt{applyabsfluxcorr=yes}, which allows for a correction of the order of 6--8\% between 3 and 12\,keV to reduce differences in the spectral shape between \textit{XMM-Newton}-pn and \textit{NuSTAR} spectra (F.\ Fürst 2022, XMM-CCF-REL-388, XMM-SOC-CAL-TN-0230).\footnote{https://xmmweb.esac.esa.int/docs/documents/CAL-SRN-0388-1-4.pdf} Finally, the background-corrected pn spectra were binned to have a signal-to-noise ratio greater than four in each spectral channel. 

The 2018 RGS spectra were also reprocessed and analysed. The individual RGS\,1 and RGS\,2 spectra at each epoch were combined into a single merged spectrum using the \texttt{SAS} task \texttt{rgscombine}. This resulted in a total count rate from the combined RGS \citep{denHerder01} spectrum of 0.273$\pm$0.002\,cts\,s$^{-1}$ over an exposure time of 51\,ks for May 2018 and 0.312$\pm$0.002\,cts\,s$^{-1}$ over an exposure time of 52\,ks for November 2018. The spectra were binned to sample the RGS spectral resolution by adopting a constant wavelength binning of $\Delta\lambda$=0.05\,\AA\ per spectral bin over the wavelength range from 6 to 36\,\AA. This ensured a minimum signal-to-noise ratio per bin of ten over the RGS energy range, allowing for $\chi^2$ minimisation in the spectral fitting procedure. 

The 2018 UV data from the \textit{XMM-Newton} Optical-UV Monitor (hereafter OM; \citealt{Mason01}) at each epoch were processed using the \texttt{SAS} script \texttt{omichain}. This script handles all necessary calibration processes (e.g., flat fielding) and runs a source detection algorithm before performing aperture photometry (using an extraction radius of 5.7$\arcsec$) on each detected source. Then it combines the source lists from separate exposures into a single master list to compute the mean corrected count rates. We quadratically added a representative systematic error of 1.5\%\footnote{https://xmmweb.esac.esa.int/docs/documents/CAL-SRN-0378-1-1.pdf} to the statistical error of the count rate to account for the OM calibration uncertainty in the conversion factor between the count rate and the flux, following the method used in \cite{Porquet19,Porquet24a,Porquet24b}.

\subsubsection{{\sl NuSTAR} Data Reduction}\label{sec:nustar}

The level 1 {\sl NuSTAR} data products were processed using the \textit{NuSTAR} Data Analysis Software (NuSTARDAS) package (version v2.1.4; released on August 22, 2024) for both focal plane modules A (FPMA) and B (FPMB). Cleaned event files (level 2 data products) were generated and calibrated using standard filtering criteria with the \texttt{nupipeline} task, along with the calibration files from the \textit{ NuSTAR} calibration database (CALDB: 20240826). For the three {\sl NuSTAR} observations and each module, source and background spectra were extracted with a radius of 60${\arcsec}$. The corresponding net exposure times are reported in Table~\ref{tab:log}. The default RMF and ARF files provided by the pipeline are on a linear energy grid with 40\,eV steps. Since the full width at half maximum (FWHM) energy resolution of \textit{NuSTAR} is 400\,eV below $\sim$50\,keV and increases to 1\,keV at 86\,keV, we rebinned the RMF and ARF files both in energy and channel space by a factor of four. This rebinning oversamples the instrumental energy resolution by at least a factor of 2.5.
 The background-corrected \textit{NuSTAR} spectra were then binned to ensure a signal-to-noise ratio greater than four in each spectral channel. To account for cross-calibration uncertainties between the {\sl NuSTAR} spectra (FPMA and FPMB) and the {\sl XMM-Newton} pn spectra, we included a free cross-normalization factor (approximately 1.2 for these datasets) in the fit to adjust for any differences between the {\sl NuSTAR} and pn spectra.

\subsection{Spectral Analysis Method}\label{sec:method}

Spectral analysis was performed using the {\sc xspec} software package (version 12.14.1; \citealt{Arnaud96}). The Galactic column density, $N_{\rm H}^{\rm Gal}$, was fixed at 1.21$\times$10$^{20}$\,cm$^{-2}$, which includes contributions from both \ion{H}{i} (1.16$\times$10$^{20}$\,cm$^{-2}$; \citealt{HI4PI16}) and \ion{H}{ii} (5.11$\times$10$^{18}$\,cm$^{-2}$; \citealt{Willingale13}). We applied the X-ray absorption model {\sc tbabs} (version 2.3.2; \citealt{Wilms00}) using interstellar medium (ISM) elemental abundances from \cite{Wilms00} and cross-sections from \cite{Verner96}. A $\chi^{2}$ minimisation was used throughout the analysis, with errors quoted at 90\% confidence intervals for one interesting parameter (i.e., $\Delta\chi^{2}$ = 2.71). For cosmological parameters, we adopted the following default values: H$_{\rm 0}$ = 67.66\,km\,s$^{-1}$\,Mpc$^{-1}$, $\Omega_{\rm m}$ = 0.3111, and $\Omega_{\Lambda}$ = 0.6889 \citep{Planck20}.

 For all fits, we consistently verified the robustness of the final parameter values and ensured that the solution is not trapped in a local (false) minimum of the $\chi^2$ space. To do so, we: (i) initiated the fits using different sets of starting values for the parameters, and (ii) used the steppar command to explore the $\chi^2$ value across the full allowed range of each parameter. This approach has proven especially valuable for parameters expressed on a logarithmic scale, such as the ionisation parameter or the disc density for the {\sc relxill} and {\sc reflkerrd} models.

\section{Spectral analysis}\label{sec:spectral-analysis}

\begin{figure}[t!]
\includegraphics[width=0.8\columnwidth,angle=0]{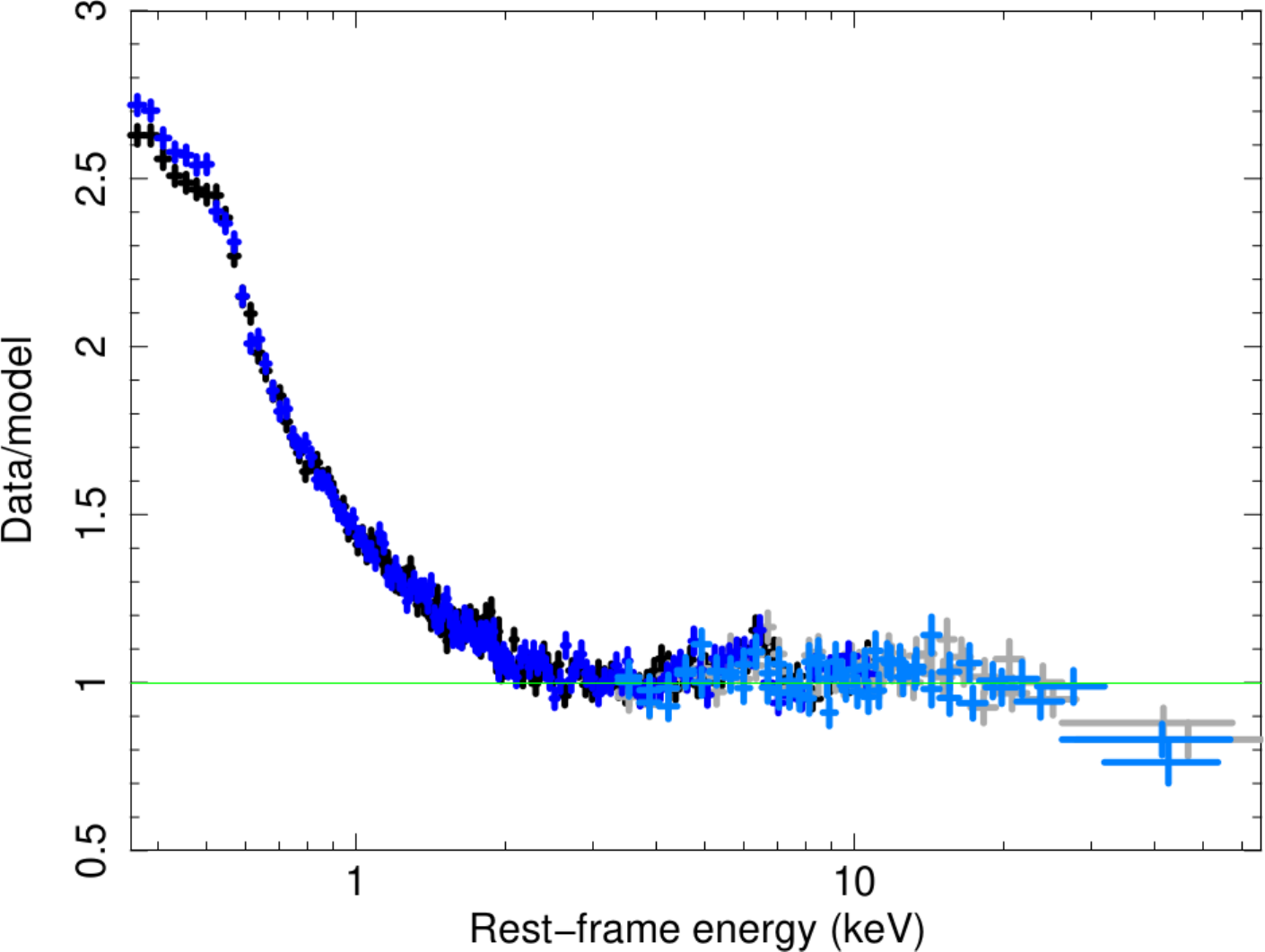}\\
	\caption{Data-to-model ratio of the two 2018 simultaneous XMM-Newton-pn and NuSTAR spectra of 1H\,0419-577 fit with a power-law model corrected for Galactic absorption over the 3--5 and 7--10 keV (AGN rest-frame) energy ranges and then extrapolated over the 0.3--79\,keV energy range. May 2018: {\sl XMM-Newton}/pn (black) and {\sl NuSTAR} (light grey). November 2018: {\sl XMM-Newton}/pn (blue) and {\sl NuSTAR} (light blue).}
\label{fig:extrapolation}
\end{figure}

\begin{figure}[t!]
\includegraphics[width=0.85\columnwidth,angle=0]{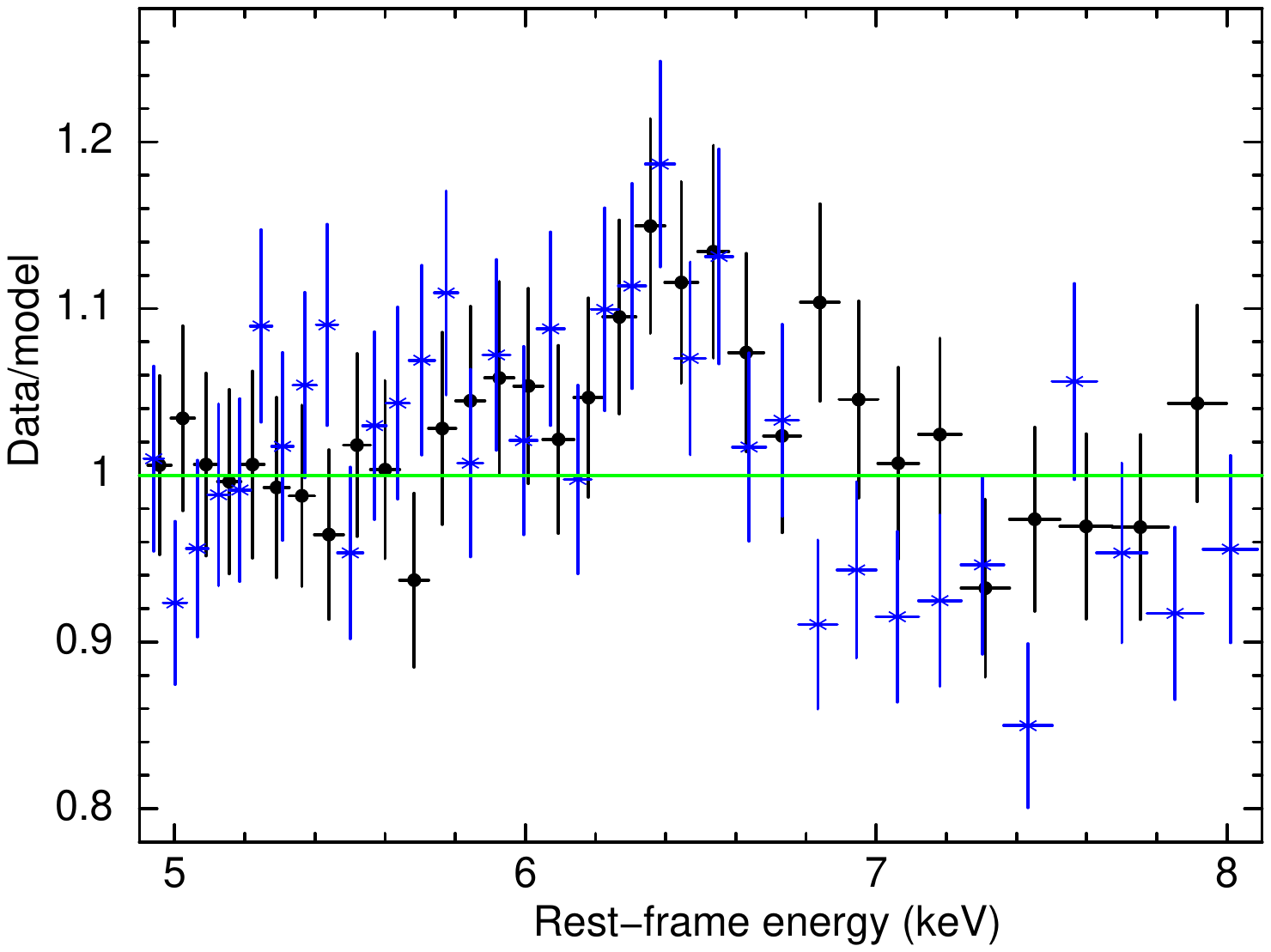}
	\caption{Zoom-in on the Fe\,K$\alpha$ line with pn data only (black: May 2018, and blue: November 2018).}
\label{fig:zoomFeK}
\end{figure}

To characterise the main X-ray components of the spectra, we simultaneously fit the {\sl XMM-Newton}-pn and {\sl NuSTAR} data over the energy ranges 3--5\,keV and 7--10\,keV (AGN rest-frame) using a power-law model with Galactic absorption. We obtained hard photon indices of $\Gamma$=1.73$\pm$0.02 and $\Gamma$=1.71$\pm$0.02 for the May and November 2018 observations, respectively. 
As shown in Fig.\ref{fig:extrapolation}, extrapolating the fit over the full energy range reveals a pronounced, smooth soft X-ray excess below 2\,keV. This excess appears to be absorption-free, with only a minor discrepancy below approximately 0.5\,keV between the two epochs. Only a weak but resolved Fe\,K$\alpha$ complex is observed (Fig.~\ref{fig:zoomFeK}). The flat hard X-ray spectrum, combined with a noticeable deficit above $\sim$30\,keV, is generally attributed to a low hot corona temperature and/or a weak (or even absent) relativistic reflection component (but see Section~\ref{sec:reflkerrd-0.3-79keV}).

\subsection{The RGS data analysis}\label{sec:RGS}

\begin{figure}[t!]
\includegraphics[width=0.9\columnwidth,angle=0]{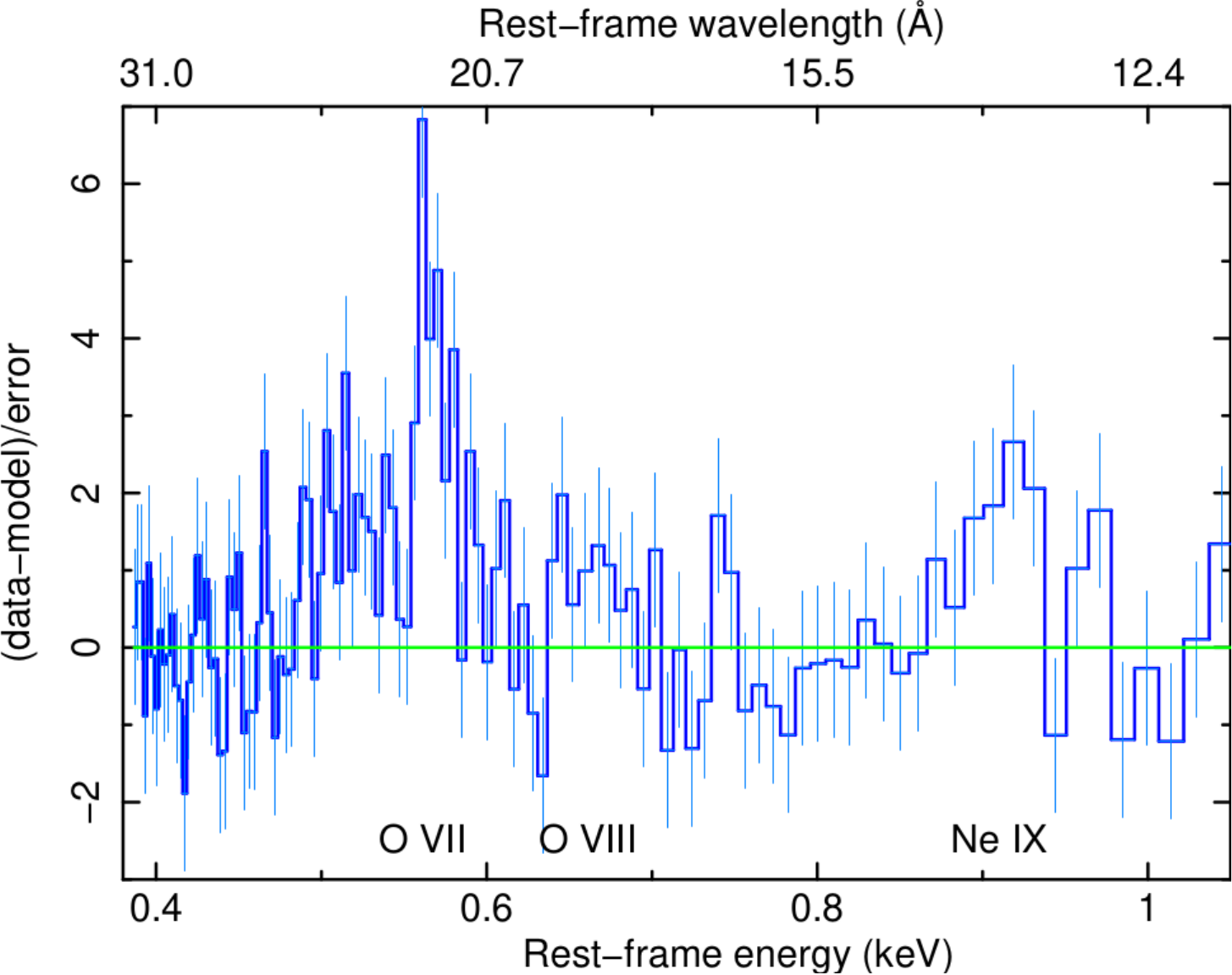}\\
\caption{Residuals from a continuum (comptt plus powerlaw) model corrected for Galactic absorption applied to the combined 2018 RGS spectrum.}
\label{fig:RGS}
\end{figure}

\begin{figure}[t!]
\includegraphics[width=0.9\columnwidth,angle=0]{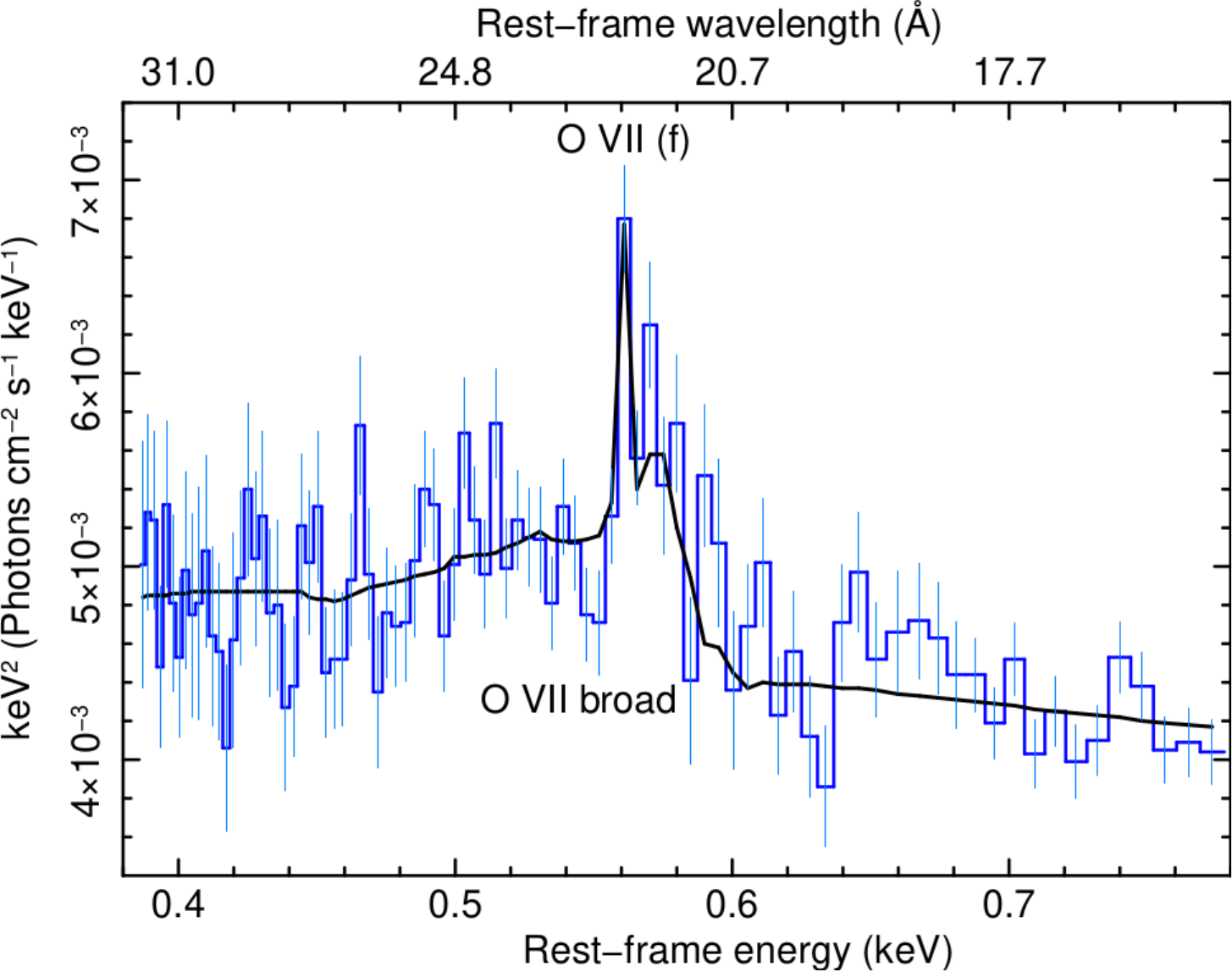}
\caption{Fit of the broad \ion{O}{vii} emission line with a relativistic line model applied to the combined 2018 RGS spectrum.}
\label{fig:OVII}
\end{figure}

Although the two 2018 X-ray broadband CCD spectra (Fig.~\ref{fig:extrapolation}) suggest that 1H\,0419-577 exhibits a similar smooth soft X-ray spectrum, our goal is to investigate any potential soft X-ray features using the high-resolution RGS spectrum. We fitted the combined 2018 RGS spectrum to characterise the underlying continuum, corrected for Galactic neutral absorption. For this, we used a combination of Comptonisation ({\sc comptt}; \citealt{Titarchuk94}) and a power-law component (with $\Gamma$ fixed at 1.7) to model the curvature of the soft X-ray spectrum. This corresponds to the baseline model: {\sc tbabs(Gal)*(comptt + powerlaw)}. The fit was found to be unsatisfactory ($\chi^{2}$/d.o.f. = 392.9/283, $\chi_{\rm red}^{2}$ = 1.39) due to the presence of several H-like and/or He-like ion emission lines around the rest-frame energies of \ion{N}{vii} Ly$\alpha$, \ion{O}{vii}, \ion{O}{viii}, \ion{Ne}{ix}, and \ion{Mg}{xi} (see Fig.~\ref{fig:RGS}).  All of these emission lines are broad, except for \ion{O}{vii}, which shows both a narrow and a broad component.

Therefore, we added six Gaussian line components to the baseline model to account for these emission features. The width of the \ion{O}{vii} narrow line was left free, while the widths of the broad lines were linked  to that of the prominent \ion{O}{vii} broad line in proportion to their energy, i.e. a constant velocity width, rather than being a constant energy width. We found that, when all broad line widths were allowed to vary independently, the resulting values were consistent with that of the broad \ion{O}{vii} line, although they were only coarsely constrained. The inclusion of these lines significantly improved the fit statistics ($\chi^{2}$/d.o.f.=262.9/273, $\chi^{2}_{\rm red}$=0.96). Specifically, adding the broad line at 576$\pm$4\,eV ($\sim$21.53\,\AA) resulted in a reduction in the $\chi^{2}$ value of  -77.5, with three degrees of freedom (Table~\ref{tab:RGSlines}). The width of the broad \ion{O}{vii} line was determined to be 17$^{+3}_{-4}$\,eV, corresponding to a FWHM of about 20\,700$^{+3\,400}_{-4\,800}$\,km\,s$^{-1}$. This value is much larger than that found for the broad-line region (BLR; FWHM(H$\beta$)$\sim$3200\,km\,s$^{-1}$), which would indicate an accretion disc origin. In contrast, the \ion{O}{vii} narrow line (FWHM$\lesssim$2100\,km\,s$^{-1}$) likely originates from a much more distant photoionised region, such as the narrow-line region (NLR). Other lines, including \ion{N}{vii} Ly$\alpha$, \ion{O}{viii} Ly$\alpha$, \ion{Ne}{ix}, and \ion{Mg}{xi}, also contributed to the improvement in the fit. The parameters of the {\sc comptt} model were determined to be $kT=0.31^{+0.21}_{-0.11}$\,keV and $\tau= 12.0^{+5.4}_{-4.1}$.\\

We then check for the presence of an ionised absorber at the Galactic ($z$=0) or the AGN rest-frame. To do this, we multiply the above model by an {\sc xstar} \citep{Kallman01} photoionised absorption table grid, constructed with an ionisation parameter, $\xi$,\footnote{$\xi$ is defined as $\equiv$$L_{\rm ion}$/($n_{\rm e}$R$^{2}$), where $L_{\rm ion}$ is the 1--1000 Rydberg ionising luminosity expressed in erg\,s$^{-1}$, $n_{\rm e}$ is the electronic density expressed in cm$^{-3}$, and $R$ is the plasma distance expressed in cm. $\xi$ is expressed in units of erg\,cm\,s$^{-1}$.} in log scale ranging from -1.5 to 3.5 and a turbulence velocity of 200\,km\,s$^{-1}$. None is required by the data, with $\Delta\chi^{2}$ of only -5.6 for two additional parameters and -3.1 for three additional parameters for a Galactic or an AGN rest-frame ionised absorber, respectively. There is no noticeable impact on the soft X-ray emission line characteristics (energies, widths, EWs). For a warm absorber (WA) in the rest-frame AGN, we obtained $N_{\rm H}^{\rm WA}\leq$1.9$\times$10$^{20}$\,cm$^{-2}$, log\,$\xi^{\rm WA}$=1.8$\pm$0.4, and v$_{\rm out}\sim$0.01\,c. This shows that any WA is negligible and points to a bare-like state of 1H\,0419-577 during these two observations.\\

\begin{table*}[h!]
\centering
\caption{Soft X-ray emission lines in the 1H\,0419-577 rest-frame observed in the combined 2018 RGS spectrum.}
\begin{tabular}{lccccc}
\hline
\hline
Line ID & Energy (eV) & $\lambda$ (\AA) & EW (eV) & $\sigma$ (eV) & $\Delta$$\chi^{2}$$^{(a)}$ \\
\hline
\ion{N}{vii} Ly$\alpha$ (broad)& 511$\pm$7 & 24.26$\pm$0.33  & 4.8$^{+1.5}_{-1.8}$ & (t) & $-$26.5 \\
\ion{O}{vii} (narrow) & 560$\pm$1  & 22.14$\pm$0.04 & 1.1$^{+0.4}_{-0.5}$ & $\leq$1.7 &  $-$15.1\\
\ion{O}{vii} (broad) & 576$\pm$4 & 21.53$\pm$0.15 & 10.6$^{+2.8}_{-1.9}$ & 16.9$^{+2.8}_{-3.9}$ &  $-$77.5\\
\ion{O}{viii} Ly$\alpha$  (broad) & 663$^{+13}_{-12}$ & 18.70$^{+0.34}_{-0.36}$ & 4.2$^{+2.2}_{-2.4}$ & (t) & $-$2.4 \\
\ion{Ne}{ix} (broad)& 908$\pm$14 & 13.65$\pm$0.21 & 6.9$^{+2.5}_{-2.9}$  & (t) &  $-$16.4\\
\ion{Mg}{xi} (broad) & 1313$^{+47}_{-37}$ & 9.45$^{+0.27}_{-0.33}$ & 6.5$^{+4.5}_{-4.6}$ &  (t) &  $-$5.2 \\
\hline
\end{tabular}
\flushleft
\small{\textit{Notes.} $^a$ Improvement in fit statistic upon successfully adding lines, listed from low to high energy. (t) means that the line width was linked to that of the broad \ion{O}{vii} line in proportion to its energy to have the same velocity width in km\,s$^{-1}$.}
\label{tab:RGSlines}
\end{table*}

Since the intense \ion{O}{vii} emission broad line appears to be compatible with an accretion disc origin, we replaced its Gaussian line component with a relativistic line model, {\sc kyrline} \citep{Dovciak04a}, while maintaining a Gaussian shape for the other broad but weaker emission lines for simplicity. We note that there is no longer a requirement for a broad \ion{N}{vii} component in this model; in fact, the excess flux at this wavelength is now fitted by the red wing of the \ion{O}{vii} relativistic line. The single emissivity index was fixed to three, while the accretion disc inclination, and its inner ($R_{\rm in}$) and outer ($R_{\rm out}$) radii were left free to vary. A good fit was found with $\chi^{2}$/d.o.f=269.4/272 (Fig.~\ref{fig:OVII}). We inferred a disc inclination angle of 28.7$^{+1.4}_{-1.3}$ degrees and values for $R_{\rm in}$ and $R_{\rm out}$ of 10$\pm$2\,$R_{\rm g}$ and 50$^{+33}_{-23}$\,$R_{\rm g}$, respectively. We then modelled all the broad emission lines with a {\sc xstar} photoionised plasma emission model.
To account for the broadening due to the apparent disc origin of the soft X-ray emission lines, we convolved the {\sc xstar} grid with the {\sc kdblur} model. Only the narrow \ion{O}{vii} component line, which is likely formed at the NLR scale, is still modelled with a Gaussian line. We fixed $R_{\rm out}$ to 50\,$R_{\rm g}$ as found previously from the {\sc kyrline} model, since it is otherwise not constrained here. We found a good fit ($\chi^{2}$/d.o.f = 287.3/279) and inferred $R_{\rm in}$=7$\pm$1\,$R_{\rm g}$, $\theta$=30$\pm$1\,degrees, $N_{\rm H}^{\rm emis}$$>$ 4$\times$10$^{22}$\,cm$^{-2}$, and log\,$\xi^{\rm emis}$=1.8$\pm$1. For a 1--1000 Ryd ionising luminosity of about 5.8$\times$10$^{45}$\,erg\,s$^{-1}$ (inferred from the SED fit of the May 2018 observation; Sect.~\ref{sec:SED}), for a radius of 10\,$R_{\rm g}$, this corresponds to a density of about 2.4$\times$10$^{15}$\,cm$^{-3}$. This value is compatible with an accretion disc origin.

\subsection{Spectral analysis above 3\,keV at three epochs: June 2015, May 2018, and November 2018}\label{sec:above3keV}

\begin{figure}[t!]
\begin{tabular}{c}
\includegraphics[width=0.85\columnwidth,angle=0]{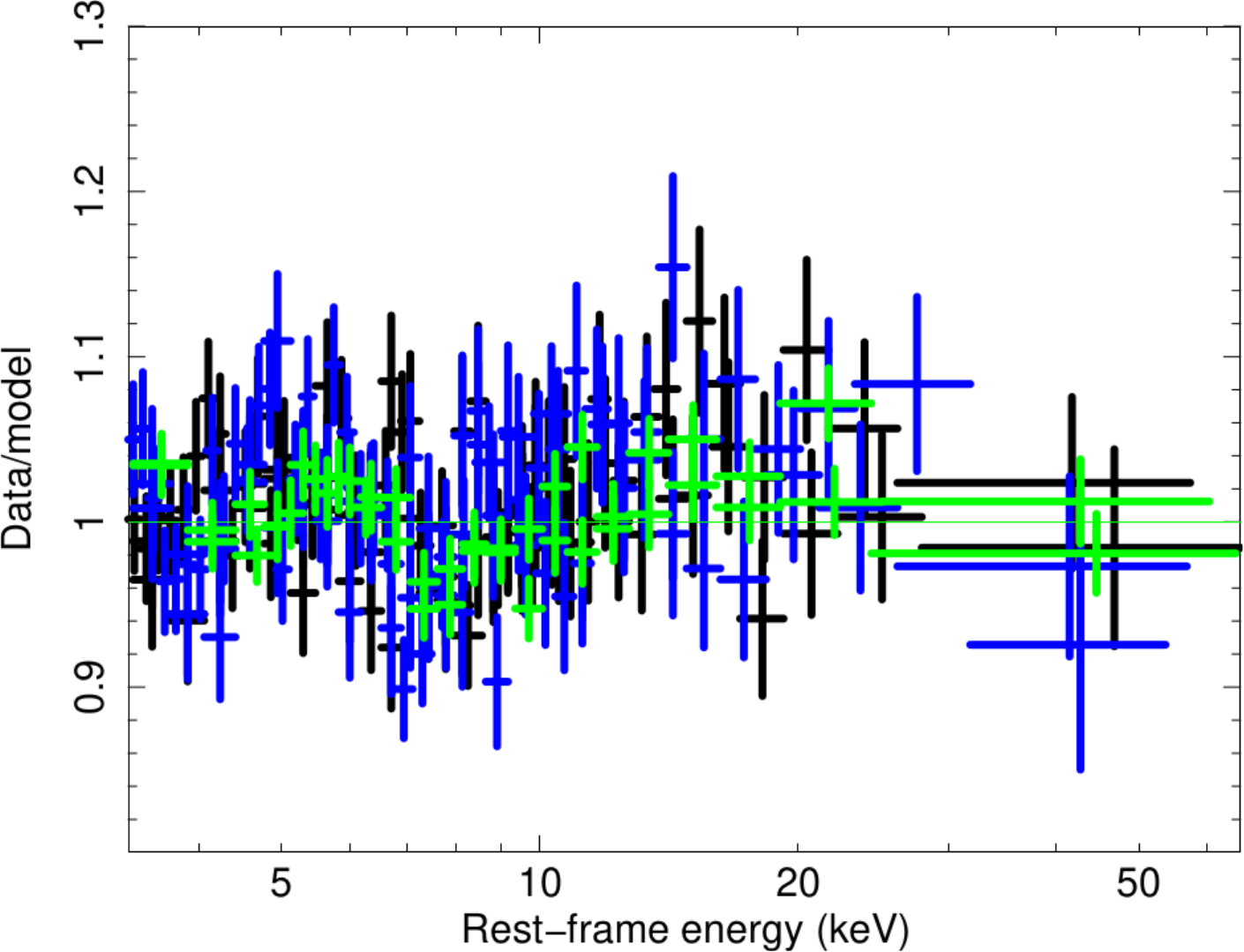} \\
\includegraphics[width=0.85\columnwidth,angle=0]{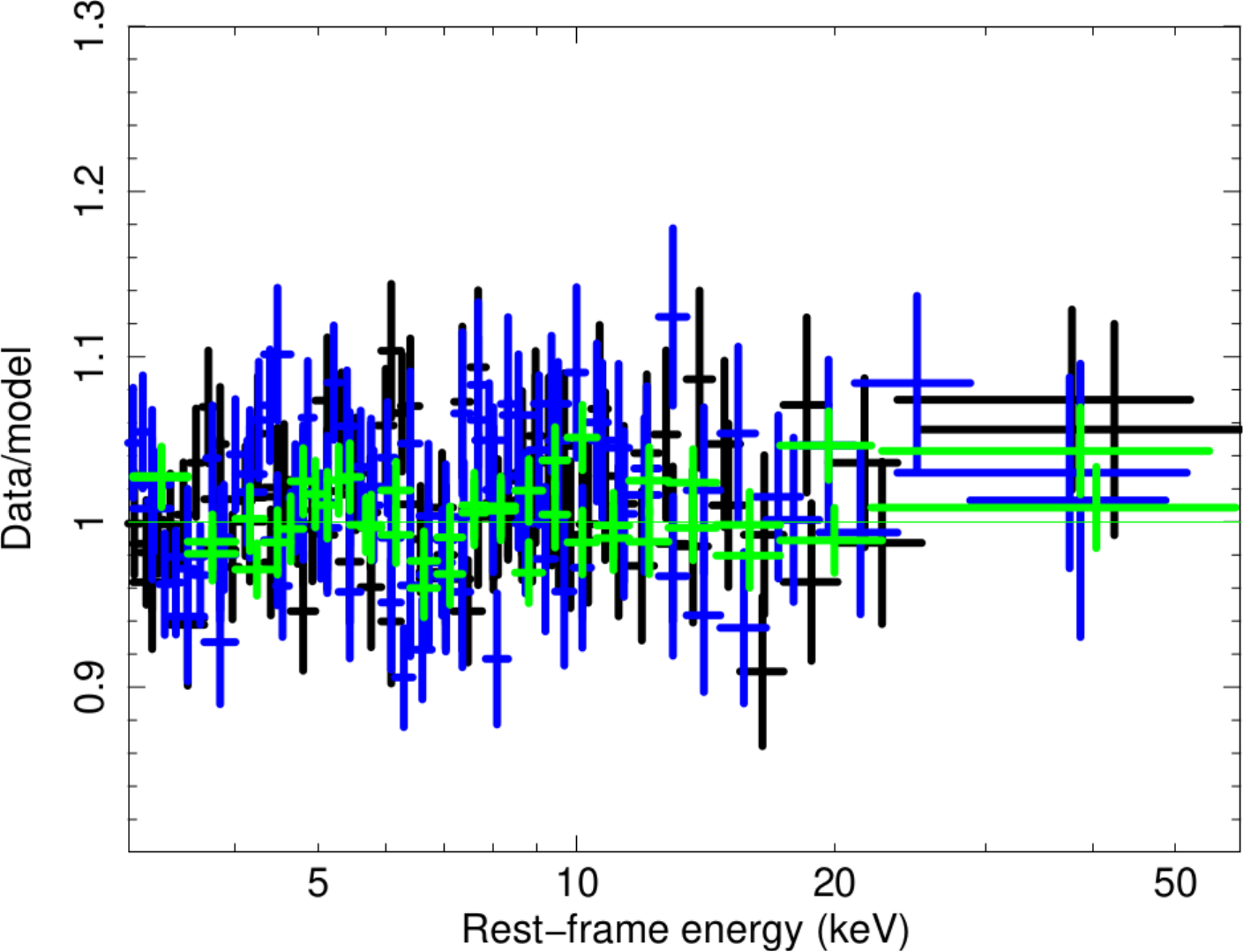}\\
\includegraphics[width=0.85\columnwidth,angle=0]{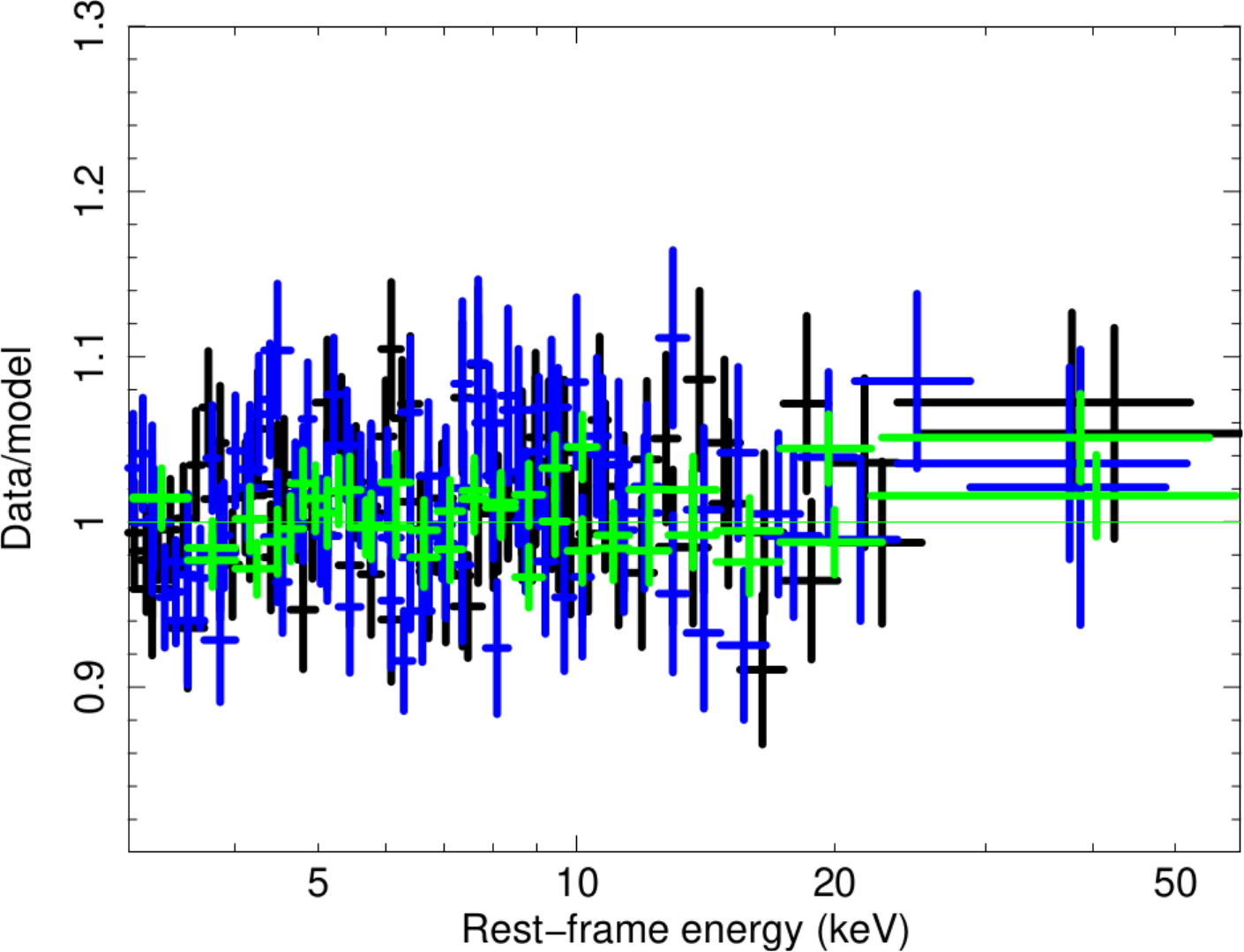}\\
\end{tabular}
\caption{Fits above 3\,keV of the two 2018 simultaneous {\sl XMM-Newton}/pn and {\sl NuSTAR} spectra and of the 2015 {\sl NuSTAR} spectra. The corresponding fit parameters are reported in Table~\ref{tab:reflabove3keV}.
Black: May 2018 {\sl XMM-Newton}/pn and {\sl NuSTAR} data.
Blue: November 2018 {\sl XMM-Newton}/pn and {\sl NuSTAR} data.
Green: June 2015 {\sl NuSTAR} data.
Top panel: {\sc zcutoffpl} model.
Middle panel: {\sc relxillcp} model.
Bottom panel: {\sc reflkerrd} model.}
\label{fig:reflabove3keV}
\end{figure}

\begin{table}[t!]
  \caption{Simultaneous fits above 3\,keV of the May and November 2018 {\sl XMM-Newton/NuSTAR} and of the June 2015 {\sl NuSTAR} observation of 1H\,0419--577. }
\centering                          
\begin{tabular}{@{}c c c c}
\hline\hline                 
 Parameters               &   \multicolumn{1}{c}{2018 May}   &   \multicolumn{1}{c}{2018 Nov} &   \multicolumn{1}{c}{2015 June} \\
                          &   \multicolumn{1}{c}{\sl NuSTAR} & \multicolumn{1}{c}{\sl NuSTAR} &   \multicolumn{1}{c}{\sl NuSTAR}\\
                          &   \multicolumn{1}{c}{+{\sl XMM}} & \multicolumn{1}{c}{+{\sl XMM}} & \\
 \hline
\hline   
& \multicolumn{3}{c}{{\sc zcutoffpl} + {\sc zga}(broad) + {\sc zga}(narrow)}\\
 \hline
 \hline
 $E_{\rm cut}$ (keV) & 84$^{+6}_{-5}$  & 96$^{+8}_{-7}$ & 64$\pm$2   \\
 $\Gamma$ &  1.62$\pm$0.01 &  1.63$\pm$0.01   &  1.58$\pm$0.01 \\
 $norm_{\rm zcutoffpl}$ ($\times$10$^{-3}$) & 2.9$\pm$0.1  & 3.4$\pm$0.1 & 4.3$\pm$1 \\
 $EW$ (narrow) (eV) &   \multicolumn{3}{c}{$\leq$12} \\
 $\sigma$ (broad) (keV) &\multicolumn{3}{c}{0.23$^{+0.10}_{-0.15}$}\\
 $EW$ (broad) (eV) & 44$^{+26}_{-18}$ & 35$^{+28}_{-19}$ & 47$^{+21}_{-15}$\\
 $\chi^{2}$/d.o.f.  ($\chi^2_{\rm red}$) & \multicolumn{3}{c}{1666.6/1546  (1.08)}\\
 \hline
 \hline
 & \multicolumn{3}{c}{{\sc relxillcp + {\sc zga}(narrow)}}\\
\hline
\hline
$kT_{\rm hot}$ (keV) &  17$^{+10}_{-3}$  &  18$^{+20}_{-3}$  & 18$^{+4}_{-2}$ \\
$\Gamma_{\rm hot}$ & 1.70$\pm$0.02  &  1.69$\pm$0.02 &  1.69$\pm$0.01\\
log\,$\xi$$^{(a)}$ &  3.8$^{+0.3}_{-0.2}$  & 3.9$^{+0.3}_{-0.2}$  & 3.9$^{+0.2}_{-0.1}$  \\
$\cal{R}$ &  0.6$^{+0.4}_{-0.2}$ & 0.6$^{+0.3}_{-0.2}$ & 0.8$^{+0.4}_{-0.3}$ \\
$norm_{\rm relxillcp}$ ($\times$10$^{-5}$) & 3.6$^{+0.7}_{-0.6}$  & 4.2$^{+0.9}_{-0.7}$ & 4.8$^{+0.9}_{-0.8}$  \\
$\chi^{2}$/d.o.f.  ($\chi^2_{\rm red}$) & \multicolumn{3}{c}{1622.7/1543 (1.05)}  \\
\hline
\hline
& \multicolumn{3}{c}{{\sc reflkerrd} + {\sc zga}(narrow)}\\
\hline
\hline
$kT_{\rm hot}$ (keV)    & 23$\pm$3    & 28$^{+7}_{-3}$         & 25$\pm$2 \\
$\tau_{\rm hot}$        & 2.6$\pm$0.1    & 2.2$\pm$0.2   & 2.4$\pm$0.1  \\
log\,$\xi$$^{(a)}$ &  3.1$\pm$0.1     & 2.7$^{+0.1}_{-0.2}$  & 2.7$^{+0.1}_{-0.2}$    \\
$\cal{R}$      & 0.40$^{+0.01}_{-0.12}$  &  0.55$^{+0.16}_{-0.04}$  & 0.50$^{+0.06}_{-0.02}$\\
$norm_{\rm reflkerr}$ ($\times$10$^{-3}$) & 2.2$\pm$0.1        & 3.0$\pm$0.1   & 4.1$\pm$0.1   \\
$\chi^{2}$/d.o.f.  ($\chi^2_{\rm red}$) & \multicolumn{3}{c}{1603.8/1543 (1.04)}  \\
\hline
\hline
$F^{\rm unabs}_{\rm 3-79\,keV}$$^{(b)}$ &  3.1$\times$10$^{-11}$ & 3.7$\times$10$^{-11}$& 3.9$\times$10$^{-11}$  \\
$L^{\rm unabs}_{\rm 3-79\,keV}$$^{(c)}$  & 9.3$\times$10$^{44}$ & 10.9$\times$10$^{44}$ & 11.7$\times$10$^{44}$\\
\hline    
\hline                  
\end{tabular}
\label{tab:reflabove3keV}
\flushleft
\small{{\bf Note}. 
$^{(a)}$ The ionisation parameter $\xi$ ($\equiv$$L_{\rm ion}$/($n_{\rm H}$R$^{2}$)) is expressed in units of erg\,cm\,s$^{-1}$.
$^{(b)}$ The 3--79\,keV unabsorbed fluxes are expressed in units of erg\,cm$^{-2}$\,s$^{-1}$.
$^{(c)}$ The 3--79\,keV unabsorbed luminosities are expressed in units of erg\,s$^{-1}$.}
\end{table}

We aim to determine the hot corona properties and the disc reflection contribution of 1H\,0419-577 during the two 2018 simultaneous \textit{ XMM-Newton} and \textit{ NuSTAR} observations. For this analysis, we considered only data above 3\,keV to avoid any influence from the soft X-ray excess emission. Additionally, we included the long 2015 {\sl NuSTAR} observation of 1H\,0419-577 to investigate any potential variability of the hot corona temperature on timescales of a few years.\\

As a first step, we fit the data of these three epochs with a phenomenological model combining a power-law ($\Gamma$) continuum with an exponential cut-off at high energy ($E_{\rm cut}$). We added two Fe\,K$\alpha$ Gaussian emission lines with a rest-frame energy fixed at 6.4\,keV: one broad (width free to vary) and one unresolved narrow (width fixed at 0\,eV) to account for any emission contribution from the accretion disc and more distant regions (BLR, NLR, and/or torus), respectively. We noticed that when the energy of the broad line was left free to vary, it was not well constrained. The normalisation of the narrow line was tied across the three epochs, as was the width (in eV) of the broad line. A satisfactory fit was obtained (Fig.~\ref{fig:reflabove3keV} top panel, Table~\ref{tab:reflabove3keV}). The equivalent widths (EWs) of the broad line were similar across the three epochs and significantly lower than the typical EWs of the broad Fe\,K$\alpha$ lines seen in unobscured AGN \citep[EW$\sim$100--150\,eV;][]{Guainazzi06,deLaCalle10,Patrick12}. The width of the broad line corresponds to a FWHM of 25\,400$^{+11\,000}_{-16\,600}$\,km\,s$^{-1}$, which is much broader -- despite the large error bars -- than that of the H$\beta$ line originating from the BLR (FWHM(H$\beta$)$\sim$3\,200\,km\,s$^{-1}$), but consistent with an origin from the inner accretion disc, as found for the \ion{O}{vii} line ($\sim$20\,700\,km\,s$^{-1}$; Sect.~\ref{sec:RGS}). The very small EW of $<$12\,eV of the narrow line suggests an insignificant contribution from any reflection in the BLR, NLR, and/or the molecular torus, as reported from previous \textit{XMM-Newton} observations \citep[e.g.][]{JiangJ19a}.

Then, we applied more physical models combining the primary Comptonisation continuum shape (hot corona) and the relativistic reflection contribution. Such a Compton continuum is more physical and exhibits a sharper high-energy roll-off compared to an exponential cut-off power-law. Furthermore, these models have the advantage of directly measuring the hot corona temperature ($kT_{\rm hot}$). We considered two models: {\sc relxillcp}\footnote{\href{https://www.sternwarte.uni-erlangen.de/~dauser/research/relxill/}{https://www.sternwarte.uni-erlangen.de/$\sim$dauser/research/relxill/}} \citep[version 2.3;][]{Dauser10,Garcia16} and {\sc reflkerrd}\footnote{The usage notes as well as the full description of the model and its associated parameters are available at \href{https://www.wfis.uni.lodz.pl/reflkerr/}{https://www.wfis.uni.lodz.pl/reflkerr/}} \citep{Niedzwiecki19}. 
For the incident Comptonisation spectrum, {\sc relxillcp} uses the {\sc nthcomp} model \citep{Zdziarski96,Zycki99}, while {\sc reflkerrd} uses the {\sc compps} model \citep{Poutanen96}. We accounted for the weak Fe\,K$\alpha$ core as previously done, by including a narrow ($\sigma$=0\,eV) Gaussian line at 6.4\,keV. The single power-law disc emissivity index ($q$; with emissivity $\propto R^{-q}$) was set at the standard value of three. The inner radius of the accretion disc was set to the inner stable circular orbit (ISCO). The disc density value was fixed at the default value (10$^{15}$\,cm$^{-3}$), and the iron abundance was set to unity in both models.  The reflection fraction $\cal{R}$ is defined as the amount of reflection $\Omega$/(2$\pi$). The disc inclination angle was set at 30$^{\circ}$ (consistent with the RGS spectral analysis of the broad \ion{O}{vii} line), while the black hole spin was fixed at 0.998. Both models provided good fits with $kT_{\rm hot}\sim$17--18\,keV for {\sc relxillcp} and $kT_{\rm hot}\sim$23--28\,keV for {\sc reflkerrd}, respectively (Table~\ref{tab:reflabove3keV}, Fig.~\ref{fig:reflabove3keV}). The values are compatible within their error bars between the two models. In addition, for each model, similar values (within their error bars) for the hot corona temperature were measured for the three epochs. 

\begin{table}[t!]
\caption{Simultaneous fits of the two 2018 {\sl XMM-Newton} and {\sl NuSTAR} spectra over the 0.3-79\,keV range with the {\sc reflkerrd} relativistic reflection model corrected for Galactic absorption (see Sect.~\ref{sec:reflkerrd-0.3-79keV} for details).}
\centering
\begin{tabular}{lccc}
\hline\hline
Parameters &  \multicolumn{2}{c}{}\\
\hline
$a$         & 0.97$\pm$0.02 & 0.90$\pm$0.01 \\
$\theta$ (degrees)  & 32.8$^{+5.8}_{-3.8}$    &  41.1$\pm$0.7  \\
$A_{\rm Fe}$      & 0.8$\pm$0.1 & 1.6$\pm$0.2 \\
disc density (log)& 15 (f) & 18.1$\pm$0.1  \\
\hline
               &           \multicolumn{2}{c}{2018 May} \\
\hline
$q_{1}$ &  $\geq$5.1 & 8.4$\pm$0.2 \\
$q_{2}$    &  2.8$\pm$0.3& 3.3$\pm$0.7 \\
$R_{\rm br}$ ($R_{\rm g}$) & 3.4$^{+2.4}_{-1.1}$ & 6.0$\pm$0.6 \\
$kT_{\rm h}$ (keV) &  80$^{+11}_{-10}$   &  420$^{+4}_{-7}$  \\
$\tau_{\rm h}$ & 0.4$\pm$0.1  & $\leq$0.05 \\
log\,$\xi$$^{(a)}$  & 0.2$^{+0.2}_{-0.1}$  & 2.8$\pm$0.1  \\
$\mathcal{R}$       &  4.5$^{+0.6}_{-0.9}$  & 0.23$\pm$0.01 \\
$norm$ ($\times$10$^{-3}$) & 3.6$\pm$0.1  &  2.6$\pm$0.1 \\
\hline
$F^{\rm unabs}_{\rm 0.3-79\,keV}$$^{(c)}$ & \multicolumn{2}{c}{4.3$\times$10$^{-11}$} \\
$L^{\rm unabs}_{\rm 0.3-79\,keV}$$^{(d)}$ & \multicolumn{2}{c}{1.6$\times$10$^{45}$} \\
\hline
\hline
              &   \multicolumn{2}{c}{2018 Nov.} \\
\hline
$q_{1}$  & $\geq$5.8 & $\geq$9.8  \\
$q_{2}$    & 3.4$^{+0.3}_{-0.2}$ & 3.1$^{+0.8}_{-0.3}$ \\
$R_{\rm br}$ ($R_{\rm g}$) &  2.4$^{+0.6}_{-0.2}$  & 5.1$^{+11.9}_{-0.2}$   \\
$kT_{\rm h}$ (keV) &  81$\pm$12 &  423$\pm$4  \\
$\tau_{\rm h}$ & 0.4$\pm$0.1 & $\leq$0.05 \\
log\,$\xi$$^{(a)}$  & 0.2$^{+0.2}_{-0.1}$ & 2.8$\pm$0.1 \\
$\mathcal{R}$ & 4.6$^{+0.6}_{-0.8}$   &  0.25$\pm$0.01  \\
$norm$ ($\times$10$^{-3}$)  & 4.1$\pm$0.1 &  2.9$\pm$0.1 \\
\hline
$F^{\rm unabs}_{\rm 0.3-79\,keV}$$^{(c)}$ & \multicolumn{2}{c}{6.1$\times$10$^{-11}$}  \\
$L^{\rm unabs}_{\rm 0.3-79\,keV}$$^{(d)}$ & \multicolumn{2}{c}{1.8$\times$10$^{45}$} \\
 \hline
 \hline
$\chi^{2}$/d.o.f.  &  1623.5/1436 & 1545.0/1437 \\
$\chi^{2}_{\rm red}$ & 1.13$^{(b)}$ & 1.08 \\
\hline
\hline
\end{tabular}
\label{tab:reflkerrd-0.3-79keV}
\flushleft
\small{{\bf Note}.
$^{(a)}$ The ionisation parameter $\xi$ is expressed in units of erg\,cm\,s$^{-1}$.
$^{(b)}$ Though the reduced $\chi^{2}$ value appears satisfactory, a prominent excess is observed on the data-to-model ratio above about 40\,keV, as illustrated in Fig.~\ref{fig:reflkerrd-0.3-79keV} (top panel).
$^{(c)}$ The 0.3--79\,keV unabsorbed fluxes are expressed in units of erg\,cm$^{-2}$\,s$^{-1}$.
$^{(d)}$ The 0.3--79\,keV unabsorbed luminosities are expressed in units of erg\,s$^{-1}$.}
\end{table}

 \begin{figure}[t!]
\begin{tabular}{c}
\includegraphics[width=0.85\columnwidth,angle=0]{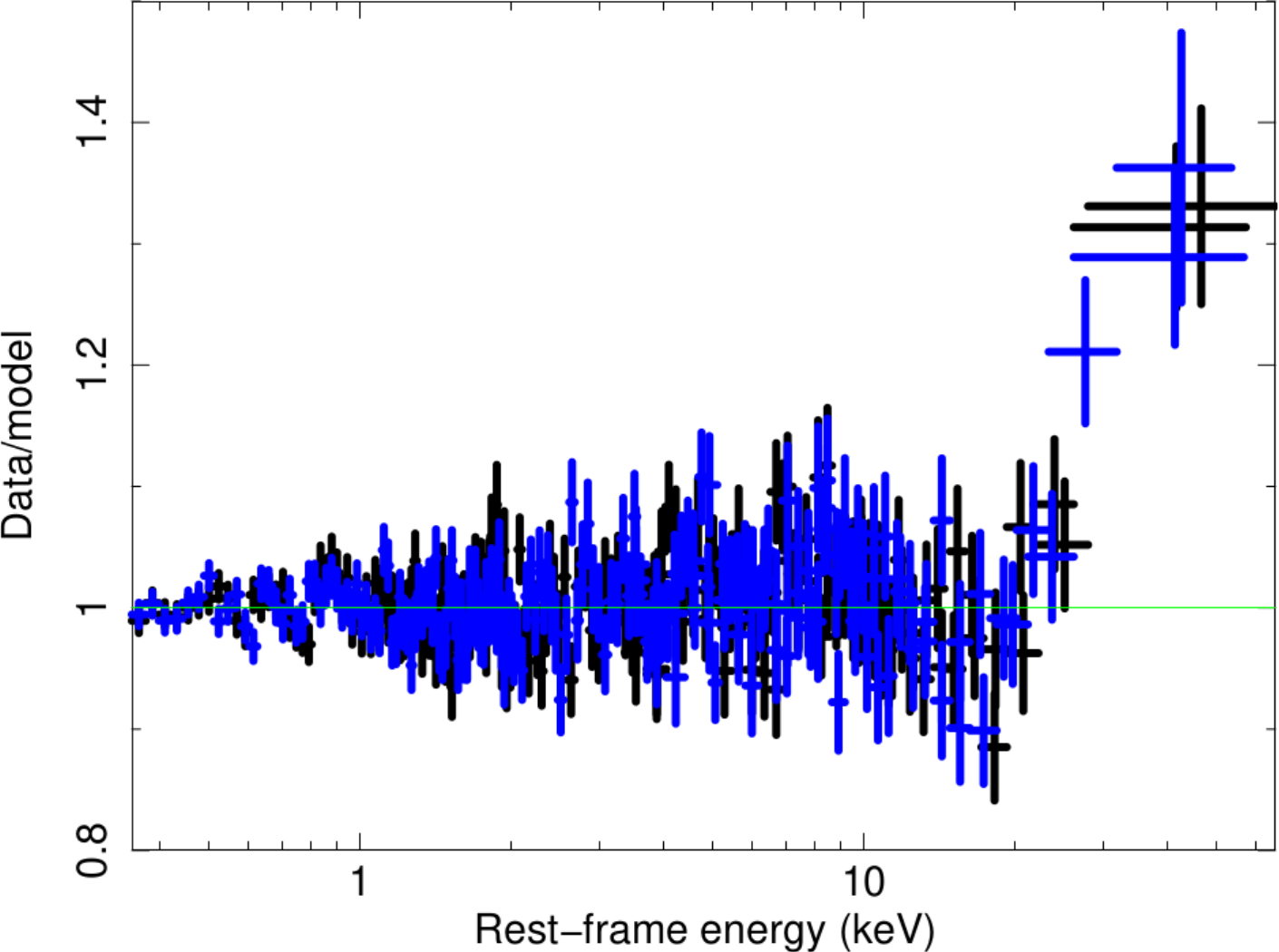}\\
\includegraphics[width=0.85\columnwidth,angle=0]{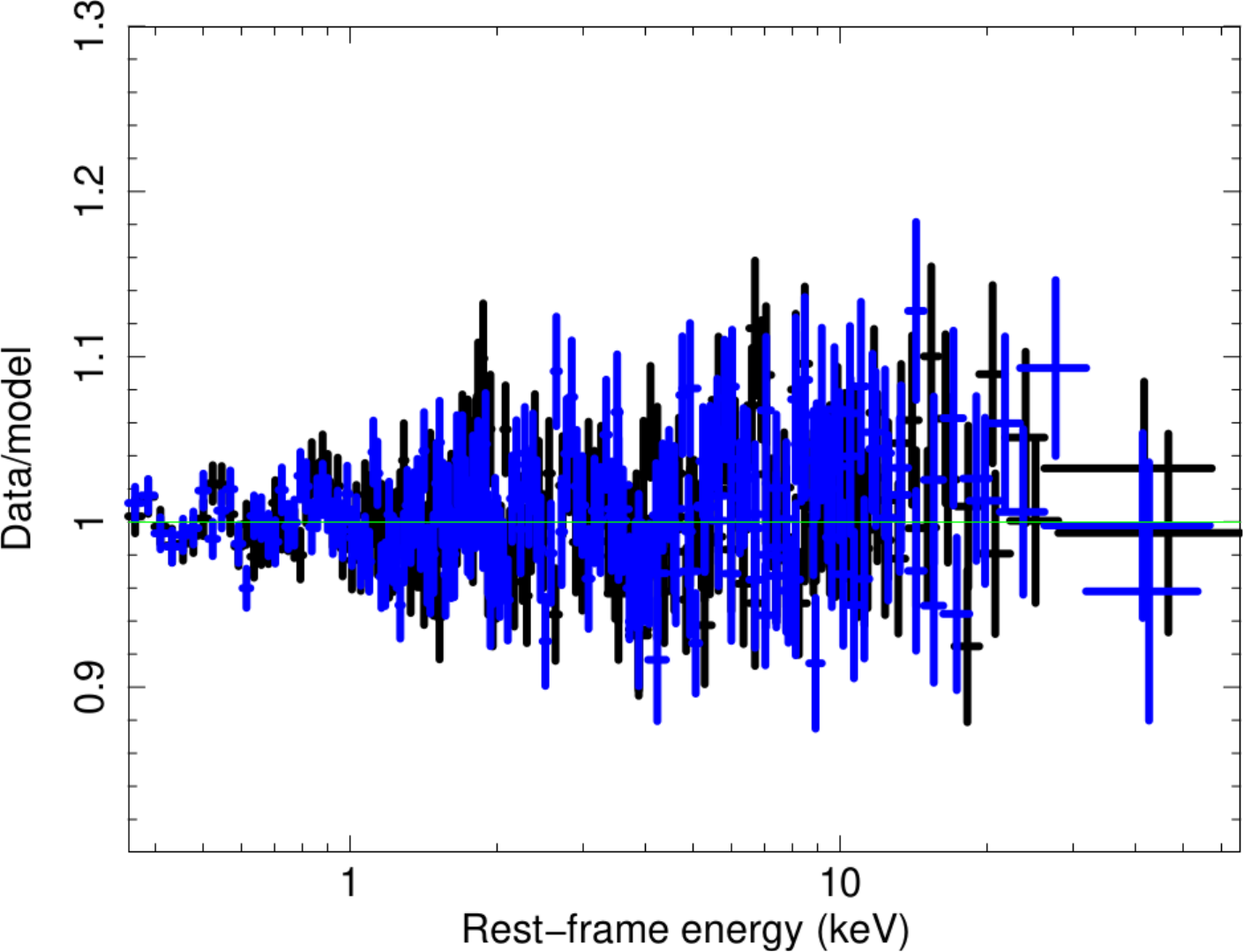}\\
\end{tabular}
\caption{Data-to-model ratio of the two simultaneous May (black) and November (blue) 2018 \textit{XMM-Newton}/pn and \textit{NuSTAR} spectra over the 0.3--79\,keV energy range, fit with the {\sc relflkerrd} model. 
Top panel: The disc density (in log scale) is set to the default value of 15. The corresponding parameter fits are reported in the second column of Table~\ref{tab:reflkerrd-0.3-79keV}.
 Bottom panel: Both the disc density and the accretion disc inclination were free to vary. The corresponding parameter fits are reported in the third column of Table~\ref{fig:reflkerrd-0.3-79keV}.}
\label{fig:reflkerrd-0.3-79keV}
\end{figure}

\subsection{Investigation of the relativistic model over the X-ray broadband range}\label{sec:reflkerrd-0.3-79keV}

We now examine the relativistic reflection model over the X-ray broadband spectra using {\sc reflkerrd} for the two simultaneous 2018 {\sl XMM-Newton} and {\sl NuSTAR} observations.  The disc density was initially set to its default value of 10$^{15}$\,cm$^{-3}$. The black hole spin and iron abundance were allowed to vary but were tied between both epochs since they are not expected to change on monthly time-scales. We allowed for a broken emissivity index for the disc at both epochs: $q_1$ is the emissivity index for $r<R_{\rm br}$, and $q_2$ is the emissivity index for $r>R_{\rm br}$, where $R_{\rm br}$ is the breaking radius expressed in $R_{\rm g}$.  A non-satisfactory fit was found over the X-ray broadband range due to a prominent positive excess above about 30 keV as a result of the steep continuum required to model the soft X-ray excess (Fig.~\ref{fig:reflkerrd-0.3-79keV}, top panel). The corresponding parameter values are reported in Table~\ref{tab:reflkerrd-0.3-79keV} (second column). A similar result was obtained when the {\sc relxillcp} model was used instead (see Appendix~\ref{app:relxillcp}). \\

However, reflection from discs with densities higher than 10$^{15}$\,cm$^{-3}$ can contribute towards the soft X-ray excess, as shown by \cite{Garcia16}. Such high-density disc models have been successfully applied to several {\sl XMM-Newton} observations of low-mass AGNs with $M_{\rm BH} \sim$ 10$^{5}$--10$^{6}$\,M${\odot}$ \citep{Mallick18,Mallick22}, and also to simultaneous or quasi-simultaneous {\sl XMM-Newton} and {\sl NuSTAR} observations of several AGNs with much higher black hole masses of $\sim$ 10$^{7}$-10$^{9}$\,M$_{\odot}$ \citep[e.g.][]{XuY21,Madathil24,Mallick25}.  
Therefore, we allowed the density of the disc to vary. For the {\sc reflkerrd} model, the disc density values span a range (in log scale) from 15 to 19. The two 2018 X-ray broadband spectra were well reproduced with a disc density value of log\,($n_{\rm e}$/cm$^{-3}$)= 18.1$\pm$0.1 (Fig.~\ref{fig:reflkerrd-0.3-79keV}, bottom panel). The best-fit parameters are reported in Table~\ref{tab:reflkerrd-0.3-79keV} (third column). We note that, when using the {\sc relxillcp} model (version 2.3) instead, where the allowed density range on a log scale is 15--20, a good fit was also found, but with a slightly higher inferred disc density of log\,($n_{\rm e}$/cm$^{-3}$)=19.5$^{+0.2}_{-0.1}$ (see Appendix~\ref{app:relxillcp}). Such a high-density disc allows for a lower reflection fraction compared to the default case (10$^{15}$\,cm$^{-3}$), due to the enhancement of the emissivity of the free-free continuum.  This leads to a consequent increase in the reflection continuum \citep{Garcia16,Ding24}. Furthermore, the disc ionisation parameter was significantly higher, with log\,$\xi$$\sim$3, compared to the model with the standard disc density value. As discussed in \cite{Ding24}, even in a highly ionised state, high-density discs can produce strong oxygen ($\sim$0.6--0.8\,keV) resonance features, as observed in the RGS spectrum (Sect.~\ref{sec:RGS}). We note that these results are not affected, even if the weak narrow Fe\,K$\alpha$ line is accounted by a more physical model, such as {\sc borus12} (Baloković et al., 2018, 2019), or if a very thin warm absorber component is included, as found by the RGS data analysis (Sect.~\ref{sec:RGS}).  Remarkably, the hot corona temperature was found to be on the order of $\sim$400\,keV, much higher than the value inferred when fitting the data above 3\,keV only ($kT_{\rm e}$$\sim$25\,keV; Sect.~\ref{sec:above3keV}). We note that if we impose the hot corona temperatures to match the values obtained from the fit above 3\,keV (Sect.~\ref{sec:above3keV}), a satisfactory fit cannot be achieved due to a large excess above $\sim$40\,keV ($\chi^{2}$/d.o.f.=1677.9/1440).
For comparison, the same modelling was performed using {\sc relxillcp}, as reported in Appendix~\ref{app:relxillcp}. As for the {\sc reflkerrd} model, much higher hot corona temperatures were found, exceeding 200--300\,keV.
However, the reflection fraction and spin values were higher than those found with {\sc reflkerrd}, while the inclination angle was much lower. This suggests that the shape of the Comptonised continuum may have a substantial impact on these parameters, at least for the present dataset. Nevertheless, despite these discrepancies, both relativistic models with a high-density disc can accurately reproduce the two 2018 X-ray broadband spectra of 1H\,0419-577.

\subsection{Investigation of hybrid model including a warm corona over the X-ray broadband and SED}\label{sec:warmcorona}

We now consider an alternative scenario in which the soft X-ray excess arises primarily from Compton up-scattering of thermal (optical/UV) photons by a population of mildly hot ($kT_{\rm e}$$\sim$0.1--1\,keV), optically thick ($\tau$$\sim$10--30) electrons in a warm corona, rather than by relativistic reflection. However, detailed analysis of the soft X-ray band -- particularly the broad \ion{O}{vii} feature -- and the emission above 3\,keV reveals clear evidence for a significant contribution from relativistic reflection in 1H\,0419-577. This necessitates a hybrid interpretation, involving emission from a warm corona, a hot corona, and relativistic reflection. Notably, in this hybrid scenario, relativistic reflection can also contribute partially to the observed soft X-ray excess.

\subsubsection{The X-ray broadband analysis}\label{sec:comptt-rexcor}

As a first step, we tested whether the soft X-ray excess could originate from a warm corona, using a simple modelling approach with the {\sc comptt} model, assuming a slab geometry. The {\sc reflkerrd} model was used to account for emission from both the hot corona and the relativistic reflection (Sect.~\ref{sec:above3keV}). For the relativistic reflection component, we assumed a single emissivity index ($q$), which was left free to vary, while the inner radius of the accretion disc was set to ISCO. The disc density was fixed at its standard value of 10$^{15}$\,cm$^{-3}$. This simple modelling provides an overall good fit to the X-ray spectra (Fig.~\ref{fig:comptt}),  suggesting that an optically thick warm corona, with $kT_{\rm warm}$$\sim$0.3\,keV and $\tau_{\rm warm}$$\sim$13,  could indeed account for the prominent soft X-ray excess in 1H\,0419-577, without requiring a high-density disc as in the pure relativistic reflection scenario (see Sect.~\ref{sec:reflkerrd-0.3-79keV}). However, the emissivity indices for both epochs are steeper than the standard value of three, indicating that significant relativistic reflection is still needed even in this warm corona scenario. This is further supported by the reflection fraction values, which are around 0.6. The best-fit parameters are reported in Table~\ref{tab:comptt}. Unlike the high-density disc scenario in the relativistic model, the inferred hot corona temperatures are comparable with those derived from the fits above 3\,keV (Sect.~\ref{sec:above3keV}).\\

We then decided to explore this hybrid scenario further using a more physically model. For this, we considered {\sc reXcor} which incorporates the effects of ionised relativistic reflection and a warm corona over the 0.3--100\,keV energy range \citep{Ballantyne20a,Ballantyne20b,Xiang22}. The accretion energy released in the inner disc is distributed among the three presumed components of the disc-corona system (Fig.\,1 in \citealt{Xiang22}): the hot corona (assumed to follow a lamppost geometry above the black hole spin axis), the warm corona, and the outer accretion disc. {\sc ReXcor} accounts for relativistic light-bending and blurring effects up to 400\,$R_{\rm g}$, using the {\sc relconv$\_$lp} convolution model \citep{Dauser13}.  Additionally, it incorporates a radially varying disc density for each annulus. Notably, when this model was applied to two {\sl XMM-Newton} observations of 1H\,0419-577 from 2010, \cite{Ballantyne24} found that the soft X-ray excess was primarily driven by emission from a warm corona rather than by relativistic reflection. Therefore, our goal is to quantitatively assess the contribution of a possible warm corona, compared to a pure relativistic reflection scenario, in shaping the soft X-ray excess of 1H\,0419-577 during the two 2018 observations.
The {\sc reXcor} fits and results are detailed in Appendix~\ref{app:ReXcor}. The inferred warm corona heating fraction values ($\sim$40--70\%, depending on the grid model) would suggest that the soft X-ray excess at both epochs is primarily dominated by warm corona emission. Intriguingly, this stands in contrast to the high-density disc scenario presented earlier, which provides an excellent fit to the X-ray broadband spectra without requiring any contribution from a warm corona component.

\subsubsection{The SED analysis}\label{sec:SED}

We now investigate the spectral energy distribution (SED), assuming a hybrid scenario, from UV to hard X-rays for the two 2018 observations, adopting a physically motivated approach to characterise the disc-corona system of 1H\,0419-577. For this, we use the new {\sc relagn} model, which builds upon the {\sc agnsed} code of \cite{Kubota18} but incorporates general relativistic ray tracing \citep{Hagen23b}. The model describes a three-zone disc-corona system: an inner hot corona ($R_{\rm ISCO} \leq R \leq R_{\rm hot}$), a warm Comptonised disc ($R_{\rm hot} \leq R \leq R_{\rm warm}$), and an outer standard accretion disc ($R_{\rm warm} \leq R \leq R_{\rm out}$). A schematic of this geometry is illustrated in Fig.~2 of \cite{Kubota18}. The parameters of {\sc relagn} are similar to those of {\sc agnsed}, with one key addition: a free parameter that accounts for a colour temperature correction to the standard outer disc ($f_{\rm col}$). In contrast, {\sc agnsed} assumes a fixed value of unity for this correction. A detailed description of the {\sc relagn} model is provided in \cite{Hagen23b}, with applications to several AGNs, including Fairall\,9, Mrk\,110, and ESO\,141-G55 \citep{Hagen23b,Mitchell23,Porquet24a,Porquet24b}. We note that when relativistic effects are not included, as for {\sc agnsed}, the accretion rate and the black hole spin may be significantly underestimated compared to {\sc relagn} \citep{Hagen23b,Porquet24a}

 \begin{table}[t!]
\caption{Best-fit result of the two simultaneous 2018 SED (UV to hard X-rays) of 1H\,0419-577 with the {\sc relagn+reflkerrd} model (see Sect.~\ref{sec:SED} for details).}
\centering
\begin{tabular}{@{}l c c}
\hline\hline
Parameters & \multicolumn{1}{c}{May 2018} & \multicolumn{1}{c}{Nov 2018}   \\
\hline\hline
$a$ & \multicolumn{2}{c}{$\geq$0.996 }\\
$\theta$ (degrees) & \multicolumn{2}{c}{21.6$^{+0.2}_{-0.6}$}\\
A$_{\rm Fe}$  &  \multicolumn{2}{c}{0.7$\pm$0.2}\\
\hline
log\,$\dot{m}$ & $-$0.20$\pm$0.01  & $-$0.26$\pm$0.01 \\
$kT_{\rm hot}$ (keV)   & 46$^{+2}_{-3}$  & 44$\pm$2  \\
$\tau_{\rm hot}$ & 0.6$\pm$0.1  &  0.7$\pm$0.1 \\
$\Gamma_{\rm hot}$ & 1.78$\pm$0.01 & 1.79$\pm$0.01\\
$R_{\rm hot}$ (R$_{\rm g}$) & 6.2$\pm$0.1 & 7.9$^{+0.1}_{-0.2}$\\
$kT_{\rm warm}$ (keV) & 0.42$^{+0.03}_{-0.02}$ & 0.43$^{+0.05}_{-0.02}$\\
$\Gamma_{\rm warm}$ & 2.50$\pm$0.02  & 2.70$^{+0.02}_{-0.01}$   \\
$R_{\rm warm}$  (R$_{\rm g}$)& 8.2$\pm$0.1 & 15.2$\pm$0.4\\
log\,$\xi$  &  0.11$^{+0.11}_{-0.05}$ & $\leq$0.10\\
$norm_{\rm reflkerrd}$ ($\times$10$^{-3}$) &  7.0$^{+0.5}_{-1.3}$ & 9.2$^{+0.3}_{-0.4}$\\
\hline
\hline
$F^{\rm unabs}_{\rm 0.001-100\,keV}$$^{(a)}$  &   2.8$\times$10$^{-10}$  &   2.3$\times$10$^{-10}$ \\
$L^{\rm unabs}_{\rm 0.001-100\,keV}$$^{(a)}$  &   8.4$\times$10$^{45}$ &   6.9$\times$10$^{45}$ \\
$\chi^{2}$/d.o.f. &  \multicolumn{2}{c}{1554.0/1445} \\
\hline    \hline
\end{tabular}
\label{tab:SED}
\flushleft
\small{{\bf Notes}. $^{(a)}$ The source fluxes (expressed in erg\,cm$^{-2}$\,s$^{-1}$) and luminosities (expressed in erg\,s$^{-1}$) are corrected from Galactic absorption and reddening.}
\end{table}

\begin{figure}[t!]
\begin{tabular}{c}
\includegraphics[width=0.85\columnwidth,angle=0]{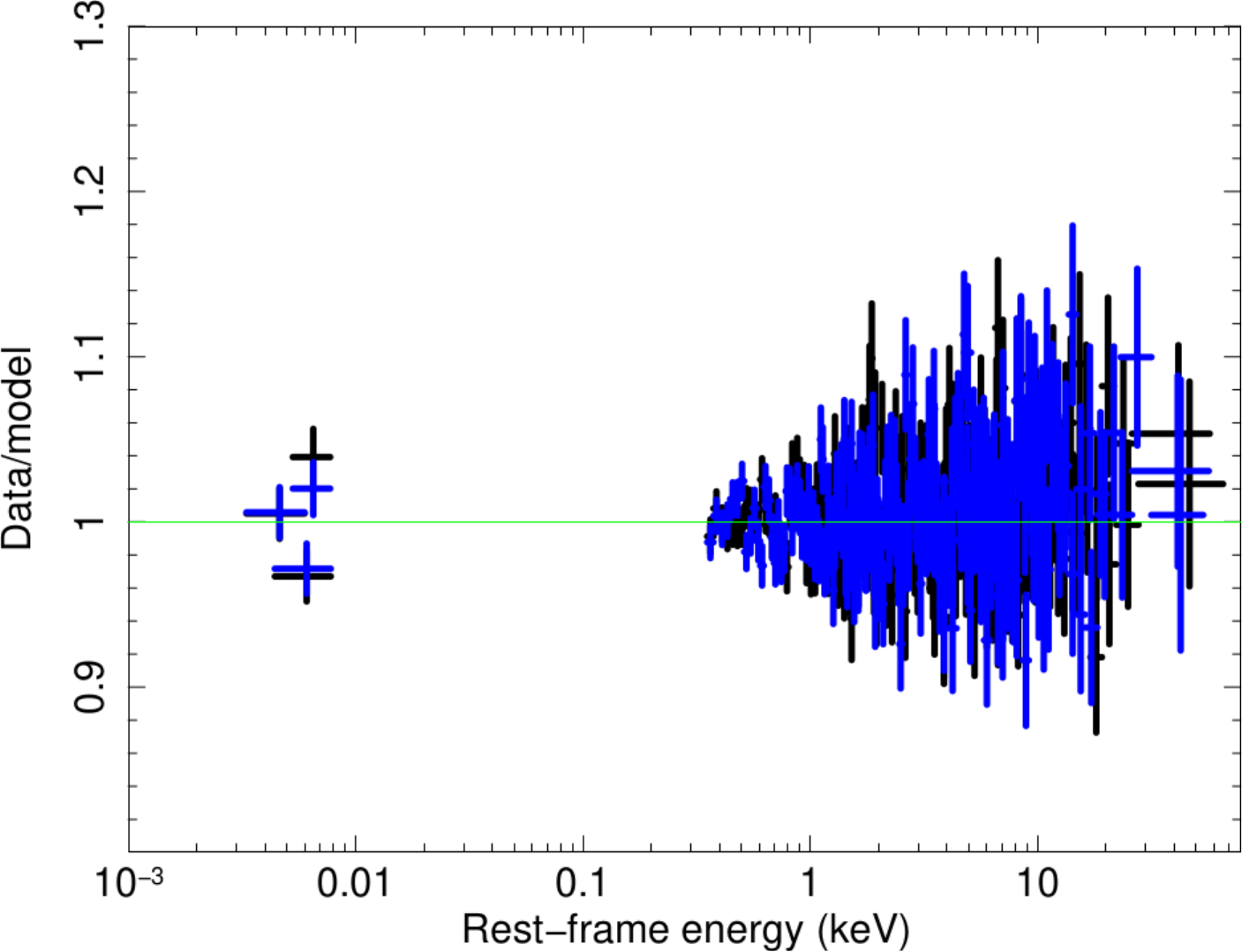} \\
\includegraphics[width=0.85\columnwidth,angle=0]{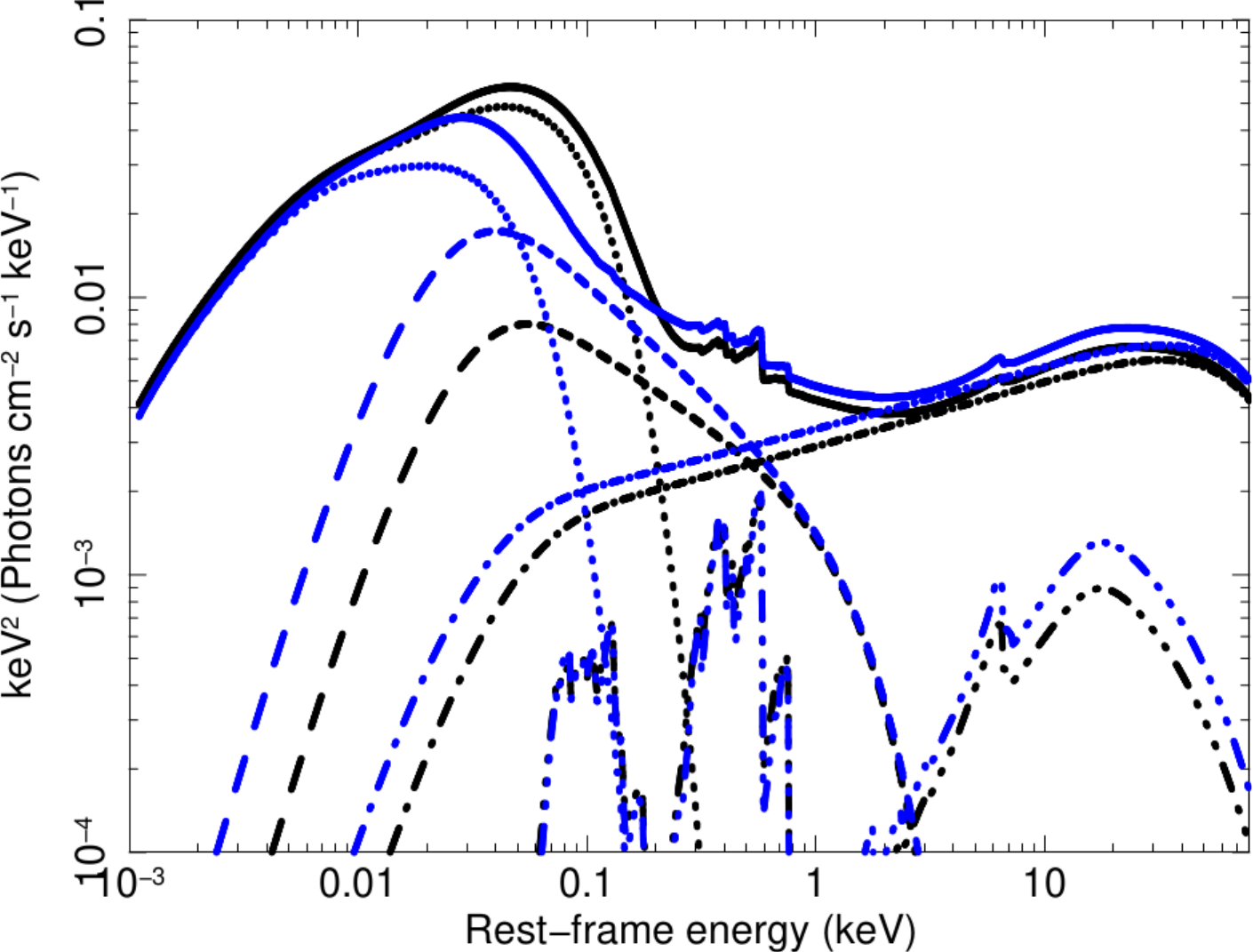}\\
\end{tabular}
\caption{SED fit from UV to hard X-rays of 1H\,0419-577 using the {\sc relagn+reflkerrd} model for the May (black) and November (blue) 2018 simultaneous {\sl XMM-Newton} and {\sl NuSTAR}. The values of the best-fit parameters are reported in Table~\ref{tab:SED}.  Top panel: Data-to-model ratio. Bottom panel: Intrinsic SED corrected for reddening and Galactic absorption (solid curves), with the main individual emission components: Outer disc (dotted curves), warm optically thick Comptonisation (dashed curves; warm corona), hot optically thin Comptonisation (dotted-dashed curves; hot corona), and relativistic reflection (three-dotted-dashed curves). For clarity, the weak narrow Gaussian line component is not displayed.}
\label{fig:SED}
\end{figure}

We used only the three shortest-wavelength UV filters from the OM instrument (UVW2, UVM2, UVW1; effective wavelengths: 2120, 2310, and 2910\,\AA, respectively), as host galaxy contamination is negligible in these bands. However, we accounted for the contribution from the broad-line region (BLR), which produces the small blue bump emission around $\sim$3000\,\AA{} \citep{Grandi82,Mehdipour15}, overlapping with the UVW1 filter. To model this contribution, we employed an additive table called {\sc smallbb}, where the normalisation is the only free parameter \citep[see][and references therein]{Petrucci18,Matzeu20}. This parameter was tied between both 2018 observations. The Galactic reddening value for 1H\,0419-577 is $E(B-V)$=0.011 \citep{Schlafly11}. For the colour correction of the outer disc, we applied the relation from \cite{Done12} by setting the $f_{\rm col}$ parameter to a negative value. The height of the hot corona ($h_{\rm max}$ in $R_{\rm g}$) was tied to its radius ($R_{\rm hot}$). We adopted a distance of 449.4\,Mpc \citep{Wright06,Planck20} and a black hole mass of 1.33$\times$10$^{8}$\,M$_{\odot}$. The inclination of the accretion disc was left as a free parameter. 

Since relativistic reflection is definitively present -- but not included in the {\sc relagn} model -- we incorporated the {\sc reflkerrd} model. The emissivity indices were both fixed at the canonical value of three. The inner radius of the relativistic reflection component was set to match $R_{\rm hot}$, as the system is truncated below this radius. The reflection fraction was fixed at -1 to account solely for the relativistic reflection component from {\sc reflkerrd}. Additionally, we included the weak narrow core of the Fe\,K$\alpha$ line using a narrow Gaussian component. Our final model is: {\sc tbabs×redden(smallbb+relagn+reflkerrd+zgaussian)}. This model provided an excellent fit from the UV to hard X-rays (Fig.~\ref{fig:SED}, top panel), with the corresponding physical parameter values reported in Table~\ref{tab:SED}. The Eddington accretion rate ratio for 1H\,0419-577 is notably high, around $\sim$0.5--0.6. The warm corona temperature, $kT_{\rm warm}$$\sim$0.4\,keV, is consistent with the value obtained using the simplified {\sc comptt} model. The moderate disc inclination angle of $\sim$22$^{\circ}$ aligns well with the value derived from the RGS data analysis of the \ion{O}{vii} broad line profile. The obtained value for $R_{\rm hot}$ ($\sim$6--7\,$R_{\rm g}$), which defines the inner radius of the disc-corona system, closely matches the inner radius of the region where relativistic reflection occurs, as inferred from the \ion{O}{vii} line profile ($\sim$10\,$R_{\rm g}$; Sect.~\ref{sec:RGS}). The warm corona appears to be rather compact, extending only about 1.3--1.9 times the size of the hot corona. From this best-fit model, we determined a lower limit of 0.996 for the black hole spin, based on the classical $\Delta\chi^2 = 2.706$ criterion. As shown in Fig.~\ref{fig:SED} (bottom panel), the warm corona emission, which in this scenario is the primary origin of the soft X-ray excess, can also significantly contribute to the UV Big Blue Bump.

\section{Summary and discussion}\label{sec:discussion}

We presented the first X-ray spectral analysis of two simultaneous {\sl XMM-Newton} and {\sl NuSTAR} observations of 1H\,0419--577, conducted in May and November 2018. The source was observed at a similarly high X-ray flux during both epochs. The broadband X-ray spectra are characterised by a prominent, absorption-free, and smooth soft X-ray excess, a weak Fe K$\alpha$ emission line, and the absence of a Compton hump.\\

From the analysis of the combined RGS spectrum of the two 2018 observations, 1H\,0419-577 can be classified as a bare-like AGN during these epochs, with an upper limit of $\sim$2$\times$10$^{20}$\,cm$^{-2}$ for the column density of any WA. However, several soft X-ray emission lines in the RGS spectrum were present, associated with the AGN and arising from He- and/or H-like ions of N, O, Ne, and Mg. Similar soft X-ray emission lines have been reported in other bright bare AGNs, such as Ark\,120 \citep{Reeves16}, Fairall\,9 \citep{Emmanoulopoulos11,Lohfink16}, and ESO\,141-G55 \citep{Porquet24b}. In these cases, the lines were attributed to emission from the BLR and/or NLR, reinforcing the idea that bare AGNs are not intrinsically devoid of gas, but rather that substantial X-ray-emitting material exists outside the direct line of sight. This interpretation applies to the narrow component of the \ion{O}{vii} emission line detected in 1H\,0419--577, which is most likely associated with a distant region such as the NLR. However, the other detected emission lines appear to be velocity broadened. The most intense broad line, \ion{O}{vii}, has a full width at half maximum (FWHM) of $\sim$20\,700\,km\,s$^{1}$, consistent with an origin in the inner accretion disc ($\sim$10--50\,R$_{\rm g}$) at an inclination angle of approximately 30$^{\circ}$, as inferred from fitting a relativistic line profile (Sect.~\ref{sec:RGS}). For comparison, the H$\beta$ line emitted by the BLR has a significantly lower FWHM of $\sim$3200\,km\,s$^{-1}$. 
Using the definition of $\xi$$\equiv$$L_{\rm ion}$/($n_{\rm e}$R$^{2}$), we inferred a density of $\sim$2.4$\times$10$^{15}$\,cm$^{-3}$, which is consistent with an origin in the accretion disc. A similarly broad \ion{O}{vii} line, originating from the inner disc, was unambiguously detected for the first time in Mrk\,110 \citep{Reeves21b}. This finding suggests that for bare AGNs with low enough Galactic column density and viewed at a moderate or nearly face-on inclination, the \ion{O}{vii} line, which is often the most prominent soft X-ray emission line, can serve as a valuable probe of relativistic reflection which may occur in the inner accretion disc, similar to the commonly used broad Fe\,K$\alpha$ line. \\

A moderately broad Fe\,K$\alpha$ line is observed, with a small EW ($\sim$40--50\,eV), which remains consistent across all three epochs. Its EW is notably lower than the typical values ($\sim$100--150\,eV) observed in unobscured AGNs \citep[see, e.g.,][]{Guainazzi06,deLaCalle10,Patrick12}. Its average width, ${\rm FWHM}$=25\,400$^{+11\,000}_{-16\,600}$\,km\,s$^{-1}$, is consistent with an origin in the inner accretion disc, in agreement with the broad soft X-ray \ion{O}{vii} emission line. The very small EW of the narrow Fe\,K$\alpha$ core ($\leq$12\,eV) indicate an insignificant reflection contribution from the BLR and/or the molecular torus, consistent with previous {\sl XMM-Newton} observations \citep[e.g.][]{JiangJ19a}.
This EW of the narrow Fe\,K$\alpha$ component in 1H\,0419-557 is significantly smaller than the typical values found for type I AGNs ($\sim$50--100\,eV; \citealt{Liu10,Shu10,Fukazawa11,Ricci14}). 
This can be attributed to a decrease in the covering factor of the molecular torus with increasing X-ray luminosity, leading to an anti-correlation between the EW of the narrow core Fe K$\alpha$ line and the X-ray continuum luminosity. This relation is known as the ‘X-ray Baldwin’ effect, or the ‘Iwasawa-Taniguchi’ effect \citep[e.g.,][]{Iwasawa93,Page04,Bianchi07,Shu10,Ricci14}. 
According to the relationship between the EW of the Fe K$\alpha$ narrow core and the 2--10 keV AGN luminosity, as established by \cite{Bianchi07}, an EW of $\sim$30\,eV is expected for 1H\,0419-577 given its 2--10\,keV luminosity of about 4$\times$10$^{44}$\,erg\,s$^{-1}$. \\

Two models were applied above 3\,keV to probe the hot corona and relativistic reflection properties, allowing for a direct measurement of the hot corona temperature \citep[{\sc relxillcp} and {\sc reflkerrd};][]{Dauser10,Garcia16,Niedzwiecki19}.  Our analysis also incorporates the deep NuSTAR dataset obtained in 2015. The hot corona temperatures are compatible within their error bars between both models, and across the three epochs (Table~\ref{tab:log}): $\sim$17--18\,keV and $kT_{\rm hot}$\,$\sim$23--28\,keV for {\sc relxillcp} and {\sc reflkerrd}, respectively. These measurements are in good agreement with previous reports for 1H\,0419-577 using \textit{NuSTAR} and/or \textit{Neil Gehrels Swift} observations \citep{TurnerJ18,JiangJ19a,Akylas21,Kamraj22,Kang22,Pal24,Serafinelli24}. We note that the measurement of the mean value of the hot corona is model-dependent as shown in Sect.~\ref{sec:above3keV}, and is often inferred in the literature from the cut-off energy value. Therefore, any robust comparison should be made under the same modelling assumption. The lack of variability of the hot corona temperature across the three epochs may be simply due to 1H\,0419-577 being observed in a similar hard X-ray flux state. Therefore, any possible ‘hotter-when-brighter’ or ‘cooler-when-brighter’ behaviour cannot be tested, as has been done for some AGNs \citep{Keek16,Ursini16,ZhangJ18,Middei19,Kang21,Pal23}.
The hot corona temperatures for 1H\,0419-577 are located in the lower range of the distribution found from AGN samples \citep[e.g., <$kT_{\rm hot}$>=50$\pm$21\,keV][]{Middei19}. Such low-to moderate hot corona temperatures have been observed in several luminous highly-accreting AGNs (e.g. \object{GRS 1734-292}, \citealt{Tortosa17}; \object{Ark\,564}, \citealt{Kara17};  \object{PDS\,456}, \citealt{Reeves21a}; and \object{IRAS 04416+1215}, \citealt{Tortosa22}), but also in luminous AGNs accreting at much lower rates (e.g., \object{ESO 362-G18}, \citealt{XuY21}; \object{ESO 511-G030}, \citealt{Zhang23}; \object{Mrk\,110},  \citealt{Porquet24a}, and \object{HE\,1029-1401}, \citealt{Vaia24}). Several physical processes have been proposed to explain such mild temperatures of the hot corona, such as  pair production (with possible non-thermal particle contribution), strong cooling of the electrons due dense UV photon fields, and/or scattering due to the relatively high optical depth of their hot corona \citep[e.g.,][]{Fabian15,Fabian17,Kara17,Ricci18}.\\

Regarding the two 2018 X-ray broadband spectra, we found that relativistic reflection alone onto an accretion disc with a standard density fixed at 10$^{15}$\,cm$^{-3}$ can be definitively ruled out, whatever the model used ({\sc reflkerrd} or {\sc relxillcp}; Sect.~\ref{sec:reflkerrd-0.3-79keV} and Appendix\,\ref{app:relxillcp}).  However, letting the density of the disc vary freely, the two broadband spectra, from the soft X-ray excess to the hard X-ray shape, can be nicely reproduced for a much higher disc density of log\,($n_{\rm e}$/cm$^{-3}$)$\sim$18.1 or of log\,($n_{\rm e}$/cm$^{-3}$)$\sim$19.5 for {\sc reflkerrd} and {\sc relxillcp}, respectively. Such a high-density disc value was also found in some other AGNs using simultaneous {\sl XMM-Newton} and {\sl NuSTAR} data and applying the {\sc relxill} model, such as \object{ESO 362-G18} \citep[log\,($n_{\rm e}$/cm$^{-3}$)=18.3;][]{XuY21} and \object{RBS 1124} \citep[log\,($n_{\rm e}$/cm$^{-3}$)=19.2;][]{Madathil24}. Recently, \cite{Mallick25} inferred from a sample of eleven AGNs, with black hole mass spanning (in log scale) $\sim$5.5--9.0, that for about 70\% of the AGNs a high-disc density model can fit their X-ray broadband spectra. For the remaining AGNs of their sample, an additional warm Comptonisation component is necessary to account for the observed soft X-ray excess.

We note that the values of the disc density for 1H\,0419-577 inferred from the two models are greater than the one found from the analysis of the broad \ion{O}{vii} emission line, $\sim$2.4$\times$10$^{15}$\,cm$^{-3}$, using $\xi$$\equiv$$L_{\rm ion}$/($n_{\rm e}$R$^{2}$). However, this definition of $\xi$ is a good approximation when photoionisation and recombination processes dominate the thermal equilibrium and ionisation balance. This becomes invalid for high-density plasmas where other mechanisms, such as free–free heating and cooling, become important \citep{Garcia16}. This may explain the discrepancy between the methods to infer the disc density. In both cases, a high black hole spin is favoured with a=0.90$\pm$0.01 and a=0.988$\pm$0.004 for {\sc reflkerrd} and {\sc relxillcp}, respectively. Interestingly, the inferred hot corona temperatures are greater than a few hundred keV, which are much higher than those found using only the data above 3\,keV (Sect.~\ref{sec:above3keV}). Therefore, at least for the present case, it would mean that relativistic reflection onto a high-density disc could mimic a lack of Compton hump in the {\sl NuSTAR} energy range, which could be otherwise interpreted as a low temperature for the hot corona.\\

Considering a standard $\alpha$ accretion disc in the case where the inner region is radiation dominated, the predicted density as a function of radius is given by \citep{Svensson94}: \[ n_{\rm e} \approx 7\times10^{12} \dot{m}^{-2} r^{3/2} \left[1-\left(\frac{3}{r}\right)\right]^{-1} (1-f)^{-3}\,{\rm cm}^{-3} \]
\noindent where $r$ is in Schwarzschild units, $\dot{m}$ is the dimensionless accretion rate, defined as $\dot{m}=\dot{M} / (\eta \dot{M}_{\rm Edd}) = \dot{M}c^2 (L_{\rm Edd})^{-1}$, and $f$ is the fraction of power radiated in the corona.
The Figure~1 of \cite{Garcia16} displays, in the black hole mass-$f$ plane, the lines corresponding to different $n_{\rm e}$$\dot{m}^{2}$ values, assuming a radius of 20\,$R_{\rm g}$ and $\alpha$=0.1. For 1H\,0419-577, this corresponds to $f$$\sim$93\% or $f$$\sim$98\% for the {\sc reflkerrd} and {\sc relxillcp} modellings, respectively. This means that in the high-density disc scenario, almost all the power would be radiated in the hot corona (X-rays) for 1H\,0419-577 during these two observations. Such values are compatible with $f$ values found for RBS\,1124 \citep[$f$$\geq$90\%][]{Madathil24}, and from sample studied in \cite{Mallick25} where the median value is 0.7$^{+0.2}_{-0.4}$.
This seems to be in contradiction to the significant optical-UV emission observed from 1H\,0419-577 \citep{Brissenden87,Guainazzi98,Turner99}, and to the 0.3--79\,keV X-ray luminosity accounting for only $\sim$20\% of the bolometric luminosity. Some additional processes can enhance optical-UV emission, in addition to the standard thermal emission of the accretion disc. These processes include X-ray reprocessing by the accretion disc, the BLR and/or a disc wind (e.g. \citealt{Troyer16,Lobban18,McHardy18,Edelson19,Matthews20,Vincentelli22,Panagiotou22,Hagen23a,Lewin24}). However, they could only partially account for the strong UV-to-X-ray flux ratio observed in 1H\,0419-577, which therefore remains challenging to explain in the present case.\\

 We also considered an alternative scenario that includes the presence of a warm corona, in addition to the hot corona and relativistic reflection. In this case, the soft X-ray excess would primarily originate from this warm corona region, with a partial contribution from relativistic reflection. This hybrid model has been shown to well reproduce the X-ray broadband spectra and SEDs of several other bare AGNs (Ark\,120, \citealt{Porquet18}; Fairall\,9, \citealt{Lohfink16,Partington24}; Ton\,S180, \citealt{Matzeu20},  Mrk\,110 \citealt{Porquet21,Porquet24a}, and ESO\,141-G55, \citealt{Porquet24b}).
\cite{ChenS25} show that it can adequately reproduce the soft X-ray excess strength and shape for a sample of 59 type I AGNs observed with {\sl XMM-Newton}.
 This scenario, in which the warm corona is proposed as the primary origin of the soft X-ray excess, may explain observed cases where the soft X-ray excess closely tracks the variability of the UV/optical disc emission, while the hard X-ray power-law component shows weaker or uncorrelated variability (e.g., Mrk\,509, \citealt{Mehdipour11}; ESO\,511-G030, \citealt{Middei23}; Mrk\,841, \citealt{Mehdipour23}; Fairall\,9, \citealt{Partington24}; Mrk\,590, \citealt{Palit25}). Indeed, the warm corona contributes to both the soft X-ray excess and the UV emission. However, this behaviour poses challenges to scenarios in which the soft excess arises solely from relativistic reflection.\\

 In general, the warm corona model requires specific physical conditions to explain the soft X-ray excess observed in AGNs. Reaching an optical depth above five demands strong magnetic support, with magnetic-to-gas pressure ratios exceeding thirty \citep{Rozanska15}. A key ingredient is “anomalous internal heating,” in which a substantial fraction of the accretion power is dissipated within the corona, a process that remains poorly understood and is usually treated as a free parameter \citep{Petrucci18,Petrucci20}. Several processes have been proposed to explain the origin of such internal heating, often attributed broadly to the dissipation of accretion energy, with specific mechanisms including magnetic heating such as the magneto-rotational instability dynamo process and reconnection, viscous accretion within the corona, and the dissipation of waves \citep{Rozanska15,Gronkiewicz23,Ma25} .
Simulations show that, only with sufficient mechanical heating, warm coronae produce smooth, line-free spectra consistent with the observed soft excess \citep{Petrucci18,Petrucci20,Ballantyne20a,Ballantyne20b}. If these specific  heating or density conditions are not fulfilled, the models predict absorption features not seen in observations \citep{Garcia19,Kara25}. In this context, hard X-ray illumination, which is believed to originate from the hot corona, also plays a vital role in maintaining the gas in a enough high-ionisation state to reduce photoelectric opacity \citep{Ballantyne20a,Ballantyne20b,Petrucci20,Xiang22}. Recent efforts incorporating realistic geometries and magnetic effects further support the existence of warm, optically thick coronae \citep{Gronkiewicz23,Kawanaka24}. \\

We applied to the X-ray broadband the {\sc reXcor} model, which combines the effects of both the emission for ionised relativistic reflection and a warm corona \citep{Ballantyne20a,Ballantyne20b,Xiang22}.
The inferred warm corona heating fraction values are $\sim$40--70\%, while the hot-corona heating fractions are very low at about 2--7\%. This would point to a soft X-ray excess dominated by the emission from a warm corona rather than relativistic reflection.
The values are consistent with those found for the two 2010 {\sl XMM-Newton} observations, where 1H\,419-577 was observed at a lower X-ray flux \citep{Ballantyne24}.
 Although the {\sc reXcor} model accounts for the radial variation in disc density across each annulus, the results it yields differ notably from those obtained with the relativistic reflection model assuming a high-density disc. Indeed, the latter model can reproduce the observed broadband X-ray spectra, including the soft X-ray excess, without invoking a warm corona. Moreover, the two models predict significantly different levels of accretion power released in the hot corona, highlighting the contrasting physical assumptions underlying each model.\\

For our analysis of the UV to hard X-ray SED of 1H\,0419-577 assuming a hybrid model, we applied the {\sc relagn} model that incorporates general relativistic-ray tracing to represent a structured disc-corona system \citep{Hagen23b}. The model divides the accretion flow into an inner hot corona, a warm Comptonised disc, and an outer standard disc. We also included relativistic reflection using {\sc reflkerrd} assuming a standard disc density of 10$^{15}$\,cm$^{-3}$.
 This hybrid model nicely reproduces the SED without requiring a disc density different from the standard value, relaxing the need for a very high fraction of the power released on the hot corona, as found for the relativistic reflection-only scenario. A high Eddington accretion rate (around 0.5--0.6), a disc inclination of approximately 22$^{\circ}$, and a black hole spin greater than about 0.996 were inferred. The hot corona temperatures ($\sim$40\,keV), although slightly higher, are comparable to those found from the fits above 3\,keV (Sect.~\ref{sec:above3keV}), in contrast to the high-density disc scenario. The hot corona is found to be relatively compact with $R_{\rm hot}$$\sim$6-7\,$R_{\rm g}$. We note that in this modelling the disc is truncated below $R_{\rm hot}$. This value is similar with that obtained from the analysis of the broad \ion{O}{vii} emission line, where we found that the inner radius of the relativistic emission starts at about 10\,$R_{\rm g}$ (Sect.~\ref{sec:RGS}). The warm corona has a temperatures of 0.4\,keV at both epochs and optical depth of about 13--14, and are consistent with those measured in AGNs \citep[$kT_{\rm warm}\sim$0.1--1\,keV, $\tau_{\rm warm}$$\sim$10--30, e.g.][]{Porquet04a,Bianchi09, Petrucci18,Palit24}. A small extension of the warm corona about only 1.3--1.9 times larger than the hot corona is found, but still contributes significantly to the UV emission, in addition to the thermal emission for the outer accretion disc.\\

In summary, both scenarios, relativistic reflection onto a very high-density disc versus a hybrid model combining warm and hot coronae with relativistic reflection, can successfully reproduce the two 2018 1H\,0419-577 datasets. However, they imply very different physical conditions, for instance, in terms of disc density, temperature, accretion power released in the hot corona, and the origin of the UV emission. These discrepancies highlight how distinct physical assumptions can lead to divergent interpretations of the same dataset. Therefore, a comparative approach that incorporates multiple modelling frameworks is essential for more robustly constraining the physical conditions in AGN environments and minimising the risk of model-dependent bias. Yet distinguishing between a high-density accretion disc producing pronounced relativistic reflection and a hybrid model (combining warm and hot coronae and relativistic reflection) remains challenging with the existing data. Further work is then required to disentangle these scenarios, potentially through multi-wavelength timing-analysis, multi-epochs X-ray broadband spectra and SEDs obtained at different flux states, higher spectral resolution data and/or time-resolved spectroscopy. In conclusion, the origin of the soft X-ray excess remains an open question not only for 1H\,0419-577, but also for many AGNs. It is likely to be diverse and may depend on factors such as the accretion state, black hole mass, and system geometry.

\begin{acknowledgements}
The authors thank the anonymous referee.
The paper is based on observations obtained with the {\sl XMM-Newton}, an ESA science mission with instruments and contributions directly funded by ESA
member states and the USA (NASA). This work made use of data from the
{\sl NuSTAR} mission, a project led by the California Institute of
Technology, managed by the Jet Propulsion Laboratory, and
funded by NASA. This research has made use of the {\sl NuSTAR}
Data Analysis Software (NuSTARDAS) jointly developed by
the ASI Science Data Center and the California Institute of
Technology. This research has made use of the SIMBAD
database, operated at CDS, Strasbourg, France. This research has made use of the NASA/IPAC Extragalactic Database (NED) which is operated by the California
Institute of Technology, under contract with the National Aeronautics and Space Administration. This work was supported by the French space agency (CNES). This research has made use of the computing facilities operated by CeSAM data centre at LAM, Marseille, France.
\end{acknowledgements}

\newpage
\bibliographystyle{aa}
\bibliography{biblio}

\appendix

\section{Results of relativistic reflection modelling over the full X-ray broad-band range}\label{app:relxillcp}

In this section, we report for comparison the spectral analysis of the two simultaneous {\sl XMM-Newton} and {\sl NuSTAR} observations using the {\sc relxillcp} relativistic reflection model (Table~\ref{tab:relxillcp-0.3-79keV}, Fig.~\ref{tab:relxillcp-0.3-79keV}), instead of {\sc reflkerrd} as used in Sect.~\ref{sec:reflkerrd-0.3-79keV}.

\begin{table}[h!]
\caption{Simultaneous fits of the two 2018 {\sl XMM-Newton} and {\sl NuSTAR} spectra over the 0.3-79\,keV range with the {\sc relxillcp} relativistic reflection model corrected for Galactic absorption.}
\centering
\begin{tabular}{lcc}
\hline\hline
Parameters & \\
\hline
$a$         & $\geq$0.99 & 0.988$\pm$0.004  \\
$\theta$ (degrees)  &   30.6$^{+3.5}_{-4.5}$       &   $\leq$7.6  \\
$A_{\rm Fe}$      & 0.6$^{+0.2}_{-0.1}$  & 2.4$^{+1.0}_{-1.1}$ \\
log($n_{\rm e}$) &   15 (f)   &  19.5$^{+0.2}_{-0.1}$  \\
\hline
               &           \multicolumn{2}{c}{May 2018} \\
\hline
$q_{1}$ &  $\geq$6.2   & $\geq$9.2 \\
$q_{2}$    & 2.7$\pm$0.2 &  4.2$\pm$0.1 \\
$R_{\rm br}$ ($R_{\rm g}$) & 2.9$^{+1.0}_{-0.4}$  &  2.8$\pm$0.2 \\
$kT_{\rm hot}$ (keV) & $\geq$222   &  $\geq$237  \\
$\Gamma_{\rm hot}$ &  2.28$\pm$0.02 &  1.71$\pm$0.01  \\
log\,$\xi$$^{(a)}$  & 0.10$^{+0.10}_{-0.08}$    & 3.2$\pm$0.1 \\
$\mathcal{R}$       &  7.3$^{+1.1}_{-0.9}$  & 3.2$^{+1.4}_{-0.3}$  \\
$norm_{\rm relxillcp}$$^{(b)}$  & 4.5$\pm$0.1 & 3.3$^{+0.2}_{-0.6}$ \\
\hline
              &   \multicolumn{2}{c}{Nov. 2018} \\
\hline
$q_{1}$  & $\geq$7.6  & $\geq$8.8 \\
$q_{2}$    & 3.2$\pm$0.2  &  4.4$\pm$0.1   \\
$R_{\rm br}$ ($R_{\rm g}$) & 2.4$^{+0.7}_{-0.2}$  & 2.6$\pm$0.1 \\
$kT_{\rm hot}$ (keV) &  $\geq$203   &  $\geq$193   \\
$\Gamma_{\rm hot}$ & 2.28$^{+0.03}_{-0.02}$  & 1.71$\pm$0.01 \\
log\,$\xi$$^{(a)}$  & 0.11$^{+0.10}_{-0.03}$ & 3.2$\pm$0.1  \\
$\mathcal{R}$ &  7.7$^{+1.2}_{-1.0}$  &  2.9$^{+0.7}_{-0.6}$ \\
$norm_{\rm relxillcp}$$^{(b)}$   & 5.0$\pm$0.1 &  4.1$^{+0.3}_{-0.7}$    \\
 \hline
 \hline
$\chi^{2}$/d.o.f.  &  1631.7/1435$^{(c)}$ & 1539.1/1434  \\
$\chi^{2}_{\rm red}$ & 1.14 & 1.07  \\
\hline
\hline
\end{tabular}
\label{tab:relxillcp-0.3-79keV}
\flushleft
\flushleft
\small{{\bf Note}.
(f) means that the value has been fixed.
$^{(a)}$ The ionisation parameter $\xi$ is expressed in units of erg\,cm\,s$^{-1}$.
$^{(b)}$ The normalisation of the {\sc relxillcp} model is expressed in $\times$10$^{-5}$.
$^{(c)}$ Though the reduced $\chi^{2}$ value appears satisfactory, a prominent excess is observed on the data-to-model ratio above about 30\,keV as illustrated in Fig.~\ref{fig:relxillcp-0.3-79keV} (top panel).
}
\end{table}

 \begin{figure}[t!]
\begin{tabular}{c}
\includegraphics[width=0.85\columnwidth,angle=0]{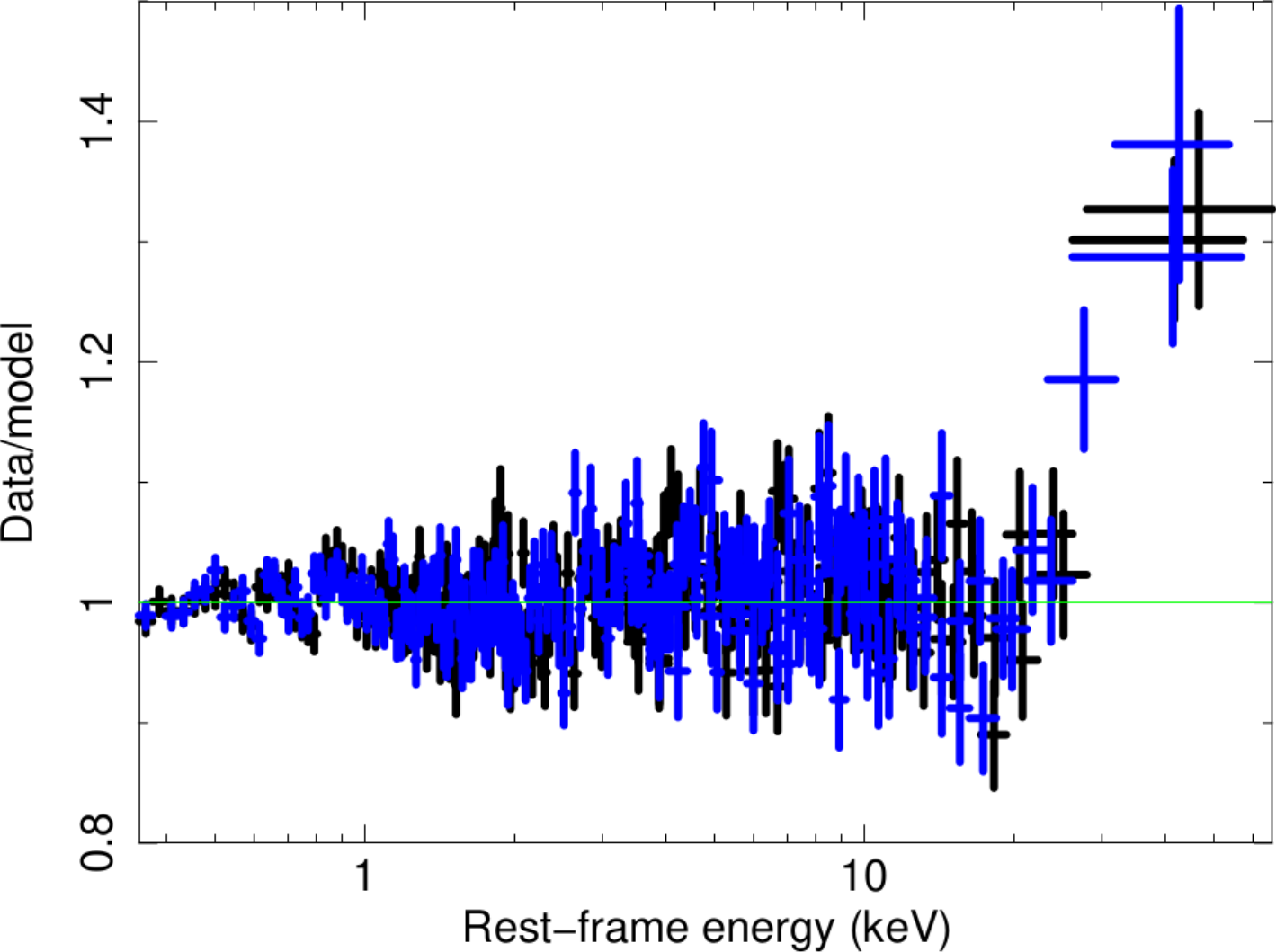}\\
\includegraphics[width=0.85\columnwidth,angle=0]{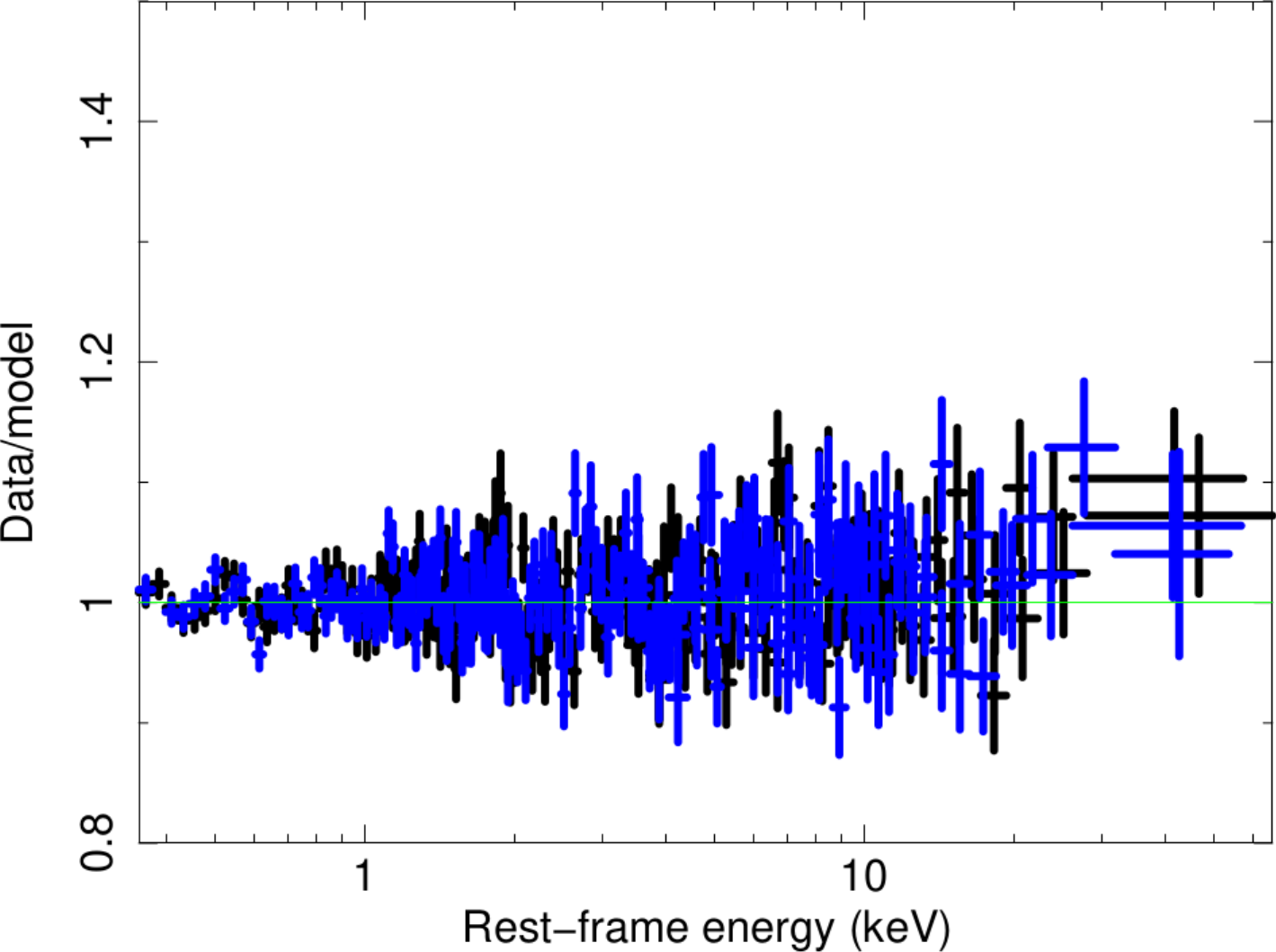}\\
\end{tabular}
\caption{Data-to-model ratio of the two simultaneous May (black) and November (blue) 2018 {\sl XMM-Newton}/pn and {\sl NuSTAR} spectra over the 0.3--79\,keV energy range fit with the {\sc relxillcp} model.
Top panel: The disc density (in log scale) was set to the default value of 15. The corresponding parameter fits are reported in Table~\ref{tab:relxillcp-0.3-79keV} second column.
 Bottom panel: The disc density and the accretion disc inclination were free to vary. The corresponding parameter fits are reported in Table~\ref{fig:relxillcp-0.3-79keV} third column.
}
\label{fig:relxillcp-0.3-79keV}
\end{figure}

\section{Results of the simple hybrid modelling combining the comptt and {reflkerrd} models}\label{app:comptt}

\begin{figure}[t!]
\includegraphics[width=0.85\columnwidth,angle=0]{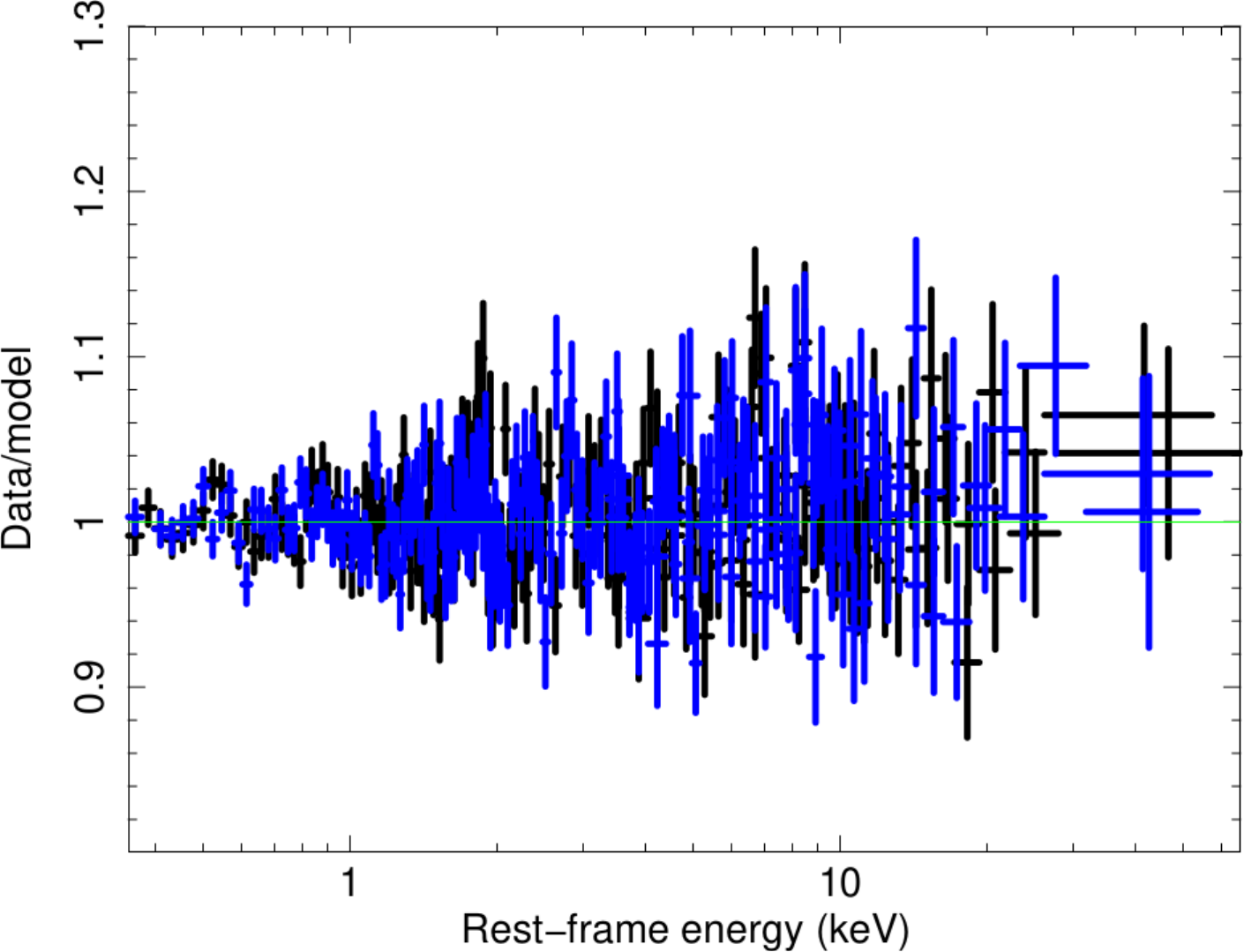} \\
\caption{Data-to-model ratio of the two 2018 0.3-79\,keV simultaneous {\sl XMM-Newton}-pn (black) and {\sl NuSTAR} (blue) spectrum of 1H\,0419-577 fit using a simplistic hybrid model combining the {\sc comptt} (warm corona) and {\sc reflkerrd} (hot corona and relativistic reflection) models (see Table~\ref{tab:comptt} and Sect.~\ref{sec:SED} for details).}
\label{fig:comptt}
\end{figure}

Table~\ref{tab:comptt} reports the fitted parameter values assuming a simple hybrid model combining the {\sc comptt} and {\sc reflkerrd} models, as discussed in Sect.~\ref{sec:comptt-rexcor}. Fig.~\ref{fig:comptt} shows the data-to-model ratio of the fit.

\begin{table}[t!]
  \caption{Fit of the two 2018 simultaneous 0.3--79\,keV {\sl XMM-Newton} and {\sl NuSTAR} observations of 1H\,0419-577 using {\sc comptt} (warm corona) and {\sc reflkerrd} (hot corona and relativistic reflection) models (see Sect.~\ref{sec:warmcorona} for details).}
\centering
\begin{tabular}{@{} c c c }
\hline\hline
 Parameters               &   2018 May     &  2018 Nov. \\

 \hline
\hline
    \multicolumn{3}{c}{{\sc comptt} (warm corona)}  \\
\hline
$kT_{\rm warm}$ (keV)                 &  0.30$^{+0.06}_{-0.04}$   & 0.27$^{+0.05}_{-0.04}$        \\
$\tau_{\rm warm}$                     &  12.9$^{+1.3}_{-1.5}$  &  13.8$^{+2.1}_{-1.6}$      \\
$norm_{\rm comptt}$  &    2.0$^{+0.3}_{-0.2}$  &   2.5$^{+0.2}_{-0.3}$ \\
\hline
 \multicolumn{3}{c}{{\sc reflkerrd} (hot corona plus relativistic reflection)} \\
\hline
$kT_{\rm hot}$ (keV)                 &  29$\pm$7     &   32$^{+7}_{-6}$       \\
$\tau_{\rm hot}$                     & 2.27$^{+0.28}_{-0.35}$  & 2.12$^{+0.29}_{-0.47}$  \\
$\theta$ (deg) & \multicolumn{2}{c}{$\leq$14.2}  \\
$a$    & \multicolumn{2}{c}{$\leq$0.83} \\
$q$           &  4.6$^{+0.6}_{-0.4}$ &   5.5$^{+0.9}_{-0.5}$               \\
log\,$\xi$$^{(a)}$        &   3.1$\pm$0.1 &   3.0$\pm$0.1   \\
$\cal{R}$        & 0.66$^{+0.24}_{-0.16}$ &  0.61$^{+0.09}_{-0.12}$\\
$A_{\rm Fe}$ & \multicolumn{2}{c}{$\leq$0.56} \\
$norm_{\rm reflkerrd}$ ($\times$10$^{-3}$) & 2.0$^{+0.2}_{-0.3}$   &  2.5$^{+0.2}_{-0.3}$     \\
\hline
$\chi^{2}$/d.o.f.  & \multicolumn{2}{c}{1521.8/1433}  \\
$\chi^{2}_{\rm red}$  & \multicolumn{2}{c}{1.06} \\
\hline
\hline
\end{tabular}
\label{tab:comptt}
\flushleft
\small{{\bf Notes}. $^{(a}$ The ionisation parameter $\xi$ is expressed in units of erg\,cm\,s$^{-1}$. $^{(a)}$.}
\end{table}

\section{ReXcor modelling}\label{app:ReXcor}

 The eight publicly available table grids of the {\sc ReXcor} models were calculated for two black hole spin values ($a$=0.90 and $a$=0.99), two hot-corona height values ($h$=5\,$R_{\rm g}$ and $h$=20\,$R_{\rm g}$), and two Eddington accretion rate values ($\dot{m}=$0.1 and $\dot{m}$=0.01). These grids used a disc inclination angle of 30 degrees and the metal abundances from \cite{Morrison83}.  The free parameters of the {\sc reXcor} grids are: the hot-corona heating fraction with 0.02$\leq$$f_{\rm X}$$\leq$0.2; the photon index of irradiating power law from the hot corona with 1.7$\leq$$\Gamma_{\rm hot}$$\leq$2.2; the warm-corona heating fraction with 0.0$\leq$ $h_{\rm f}$$\leq$0.8: $h_{\rm f}$=0 means that the soft X-ray excess is exclusively due to relativistic reflection; and the warm-corona Thomson depth with 10$\leq$$\tau_{\rm warm}$$\leq$30.

\begin{figure*}[t!]
\begin{tabular}{cc}
\includegraphics[width=0.85\columnwidth,angle=0]{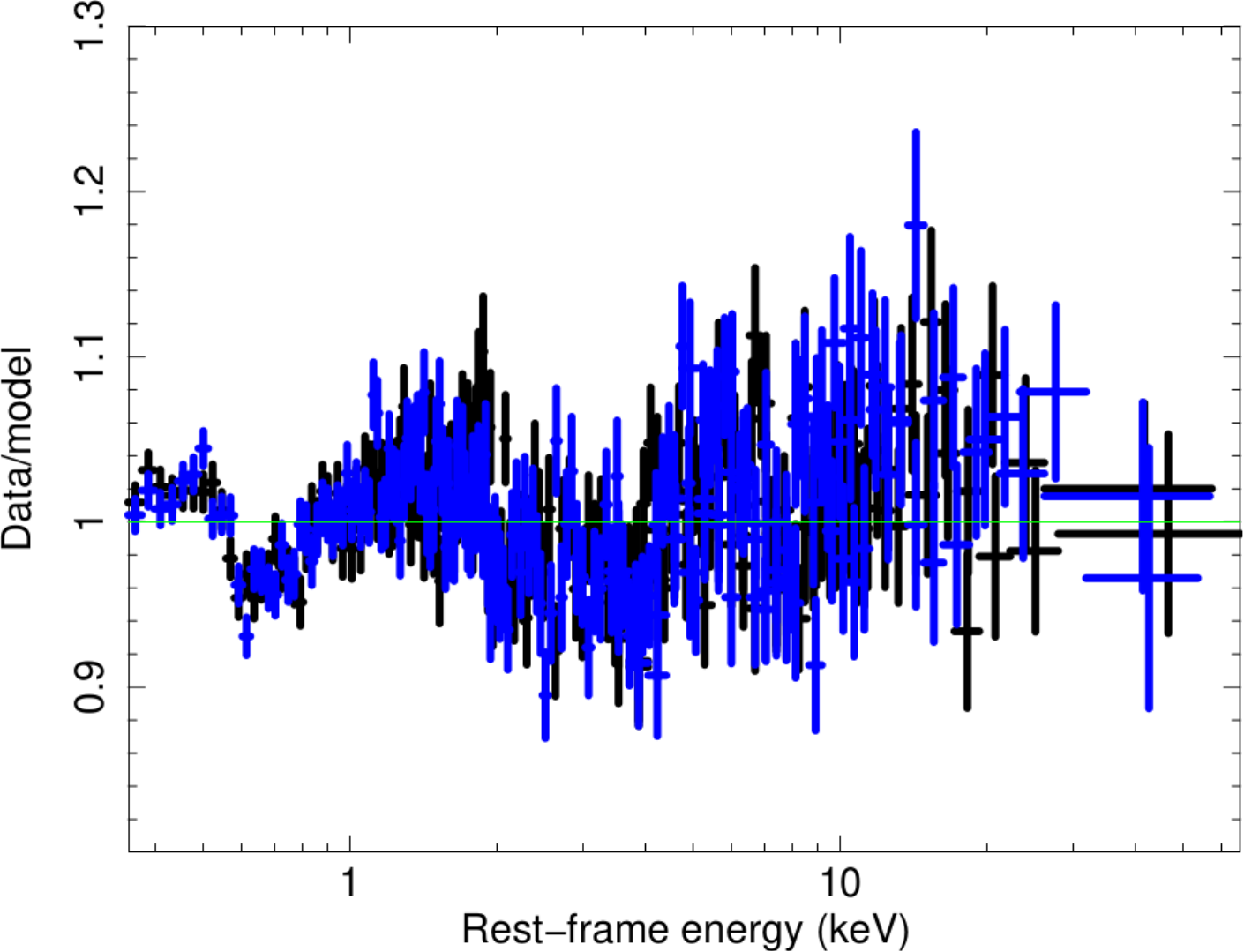} &
\includegraphics[width=0.85\columnwidth,angle=0]{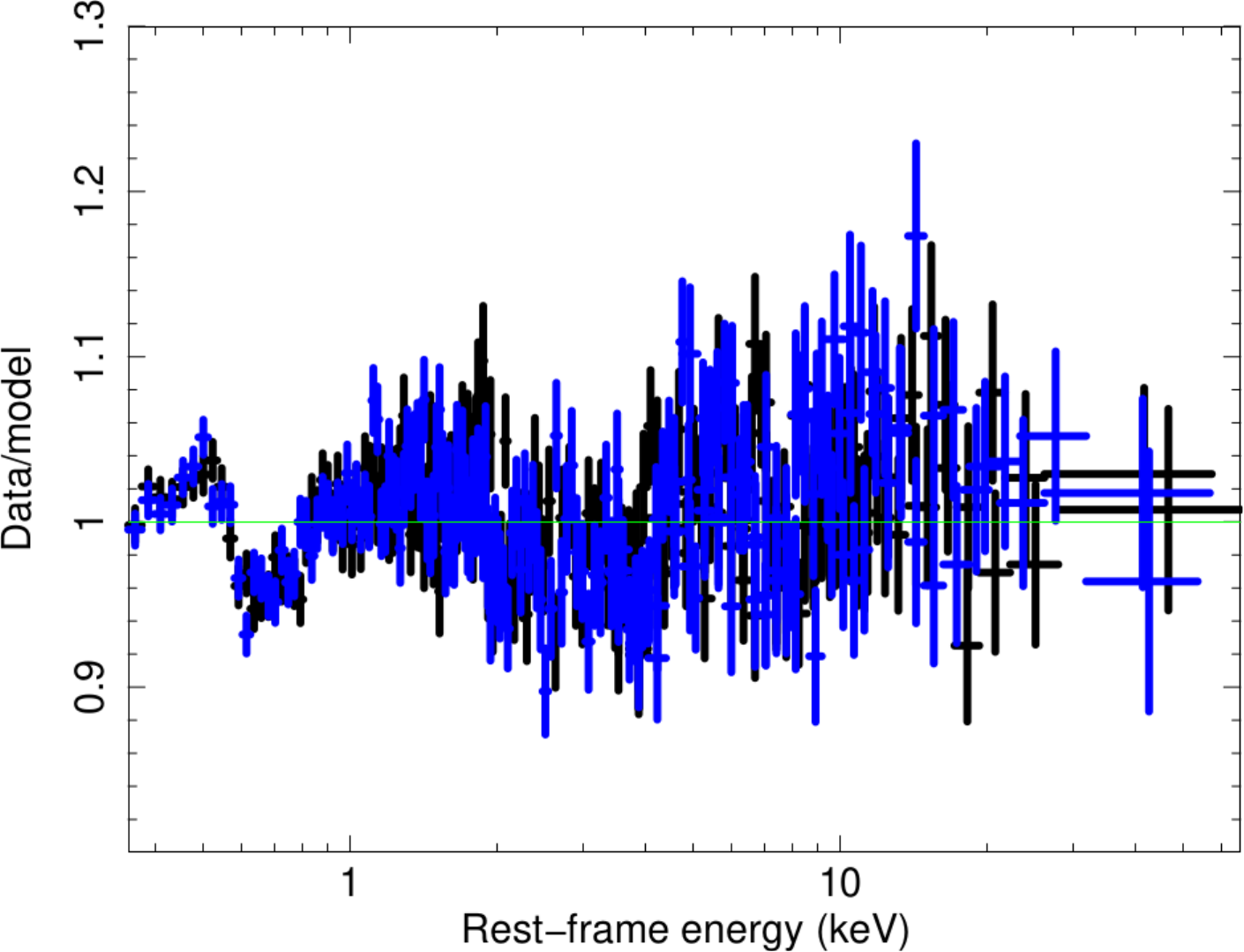} \\
\includegraphics[width=0.85\columnwidth,angle=0]{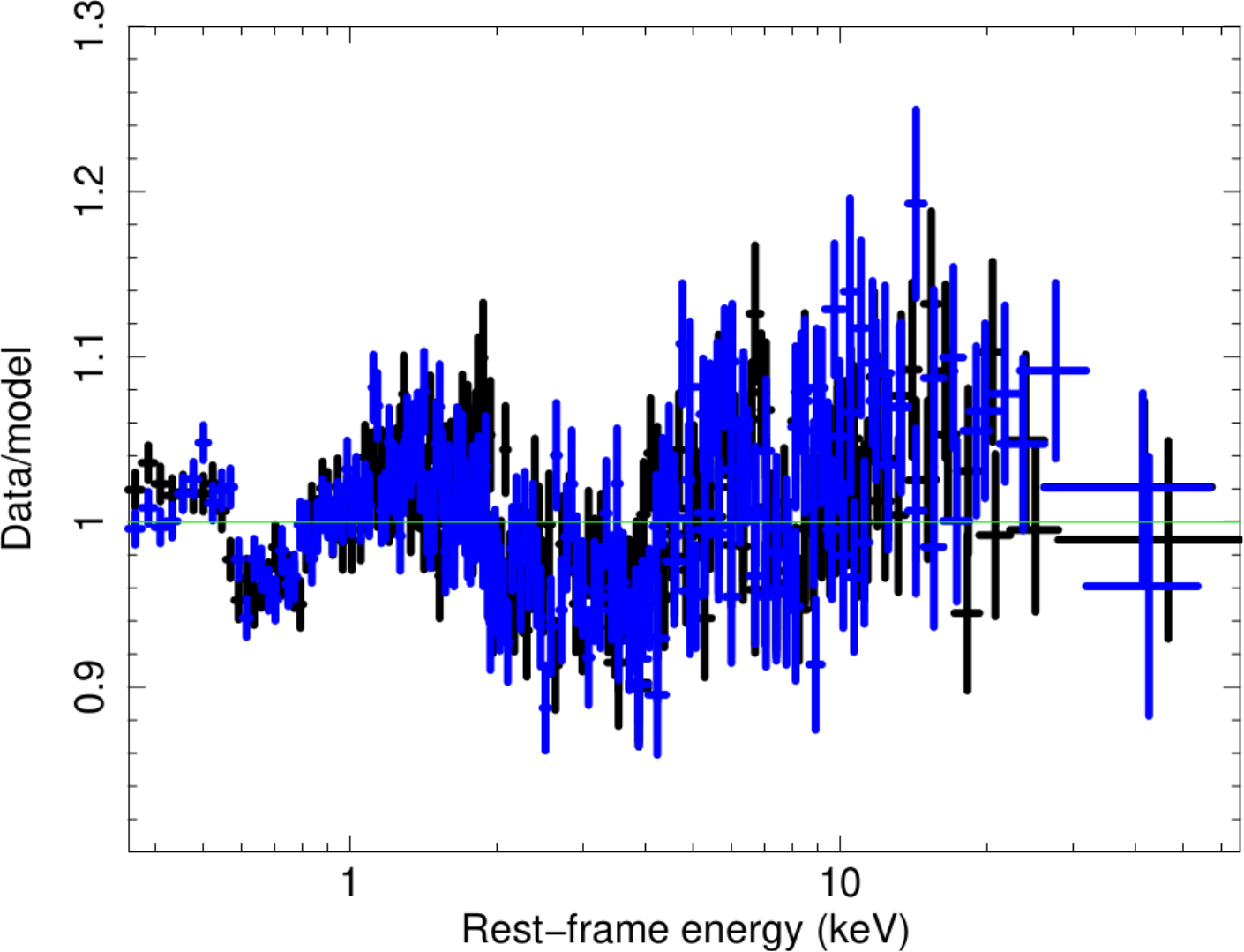} &
\includegraphics[width=0.85\columnwidth,angle=0]{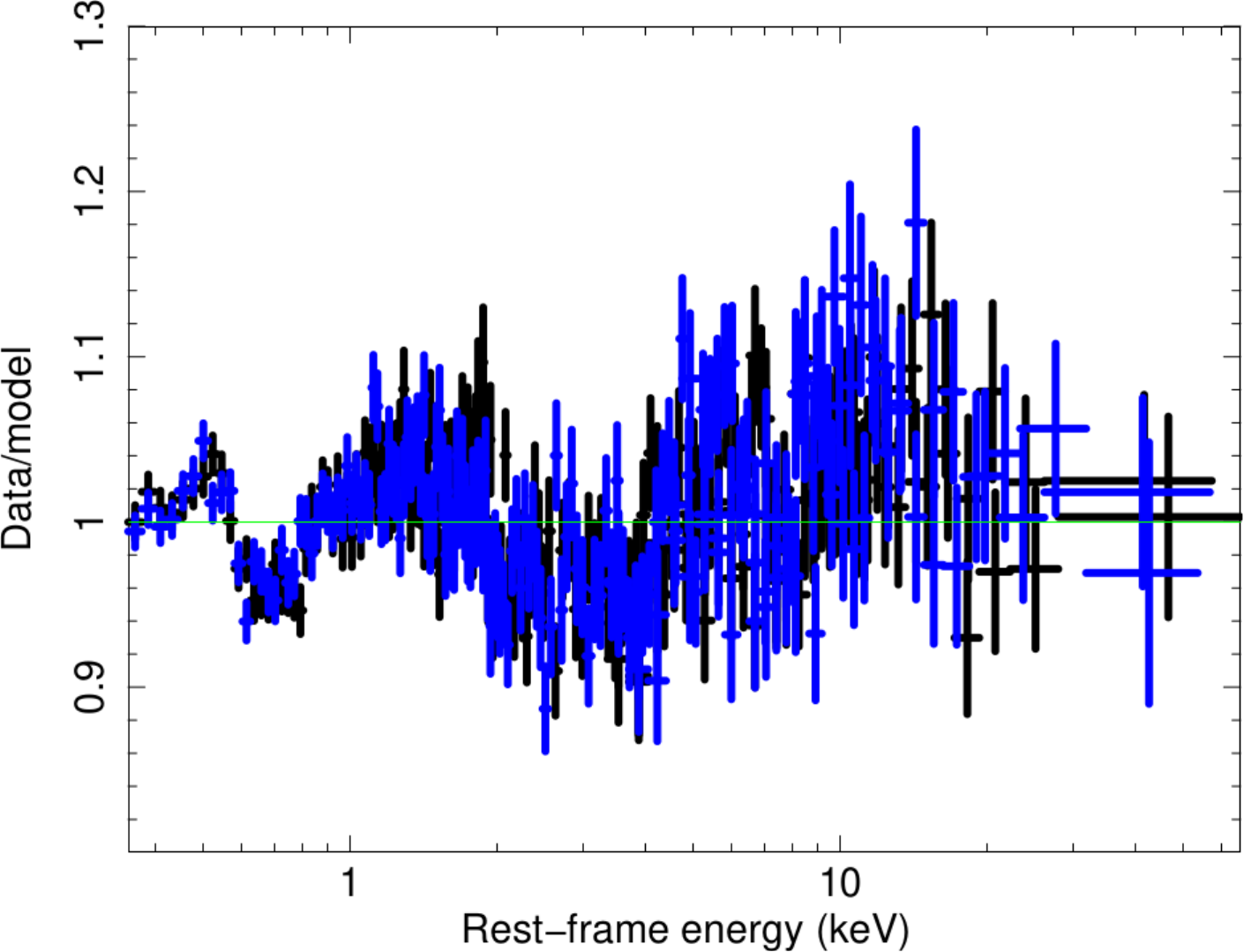} \\
\end{tabular}
\caption{Data-to-model ratio of the fits using the {\sc reXcor} model grids for the May 2018 (black) and November 2018 (blue) simultaneous {\sl XMM-Newton} and {\sl NuSTAR} spectra. The inferred parameter values are reported in Table~\ref{tab:rexcor}. Top left panel: Model calculated for a spin of 0.99 and a hot corona height of 5\,$R_{\sl g}$ ($\chi^{2}$/d.o.f.=1871.5/1437). Top right panel: Model calculated for a spin of 0.99 and a hot corona height of 20\,$R_{\sl g}$ ($\chi^{2}$/d.o.f.=1864.8/1437). Bottom left panel: Model calculated for a spin of 0.90 and a hot corona height of 5\,$R_{\sl g}$ ($\chi^{2}$/d.o.f.=1957.1/1437). Bottom right panel: Model calculated for a spin of 0.90 and a hot corona height of 20\,$R_{\sl g}$ ($\chi^{2}$/d.o.f.=1970.8/1437).}
\label{fig:rexcor}
\end{figure*}

 The {\sc reXcor} grid models do not include the underlying hard X-ray power-law continuum.  Therefore, we also added the thermally comptonised continuum model {\sc nthcomp}, allowing us to infer the hot corona temperature \citep{Zdziarski96,Zycki99}. We note that if instead a simple cutoff power-law model was applied, the hard X-ray range would not be well reproduced. The weak Fe\,K$\alpha$ narrow core is accounted for by including a Gaussian emission line ($\sigma$=0\,eV) at 6.4\,keV (Sect.~\ref{sec:above3keV}). We applied the grids with the larger Eddington accretion rate, $\dot{m}$=0.1, which is more appropriate for 1H\,0419-577 than $\dot{m}$=0.01, though its Eddington accretion rate is higher ($\sim$0.5--0.6; Sect.~\ref{sec:SED}). The model is: {\sc tbabs(Gal)$\times$(reXcor+nthcomp+zgaussian)}. The two 2018 {\sl XMM-Newton} and {\sl NuSTAR} spectra were fitted simultaneously with $h_{\rm f}$, $\Gamma$, $f_{\rm h}$, and $\tau_{\rm T}$ allowed to vary between the two epochs. We note that contrary to \cite{Ballantyne24} who analysed two previous {\sl XMM-Newton} observations, we did not include, in the modelling, any warm absorber component since, as inferred from the {\sc RGS} data analysis no significant warm absorber was observed during these two 2018 observations. 
Statistically speaking, the best-fit result was found for the model grid calculated for a spin value of 0.99 and a lamppost height of 20\,$R_{\rm g}$ (Table~\ref{tab:rexcor} and Fig.~\ref{fig:rexcor} top right panel). At both epochs, the warm-corona heating fraction values, $h_{\rm f} \sim 40$–$70\%$, suggest that a warm corona would primarily produce the soft X-ray excess. The inferred warm-corona optical depth, $\tau_{\rm warm}$$\sim$12--15, is consistent with the values obtained using the {\sc comptt} model ($\tau$$\sim$13--14; Sect.~\ref{sec:warmcorona}). Based on this modelling, only a small fraction of the accretion energy ($f_{\rm X}$$\leq$2--3\%) would be dissipated in the lamppost hot corona, contrary to the high-disc density scenario. As shown in Fig.~\ref{fig:rexcor}, the data-model ratio for all grids shows some noticeable deviations, especially in the soft X-ray range below about 1\,keV. The absorption-like feature is due to the residual of the model and not to a genuine absorption feature since none were found in the 2018 RGS data analysis (Sect.~\ref{sec:RGS}).  However, as noted by \cite{Xiang22}, the sensitivity of He-like triplets to temperature, density, and optical depth \citep{Porquet10} is not accurately described by the {\sc reXcor} model, potentially leading to residuals of a few percent. Furthermore, in this study, we used grids computed for an Eddington accretion rate of 0.1, whereas during the two 2018 observations, the accretion rate of 1H\,0419-577 was significantly higher ($\sim$0.5–0.6). This discrepancy may result in an overestimation of the \ion{O}{vii} line flux.
Furthermore, as highlighted by \cite{Ballantyne24} and \cite{Xiang22}, several limitations remain that must be taken into account. i) {\sc ReXcor} assumes a lamppost corona. Yet, for the radio-quiet type 1 AGN observed so far with {\sl IXPE}, slab-like or wedge-shaped geometries are favoured rather than a spherical lamppost or a conical shape (e.g. \object{NGC 4151}, \citealt{Gianolli23,Gianolli24}, \object{MCG-05-23-16} \citealt{Marinucci22,Tagliacozzo23}, \object{IC 4329A} \citealt{Ingram23}). ii) Currently, the model can only be applied to data above 0.3\,keV, whereas the contribution of the accretion disc (optical-UV) to power the hot corona should be included. iii) The available grids are for two Eddington accretion rates (0.01 and 0.1). In this work, we applied the four grids corresponding to the Eddington accretion rate of 0.1. However, this assumption may not be appropriate for 1H\,0419-577, which was accreting in these two 2018 observations at a significantly higher rate, $\sim$0.5--0.6.
This provides strong motivation for the extension of the {\sc ReXcor} grid's parameter ranges, to adapt it for SED analysis, as well as the possibility of incorporating a slab-like hot corona.

 \begin{table}[t!]
\caption{Best-fit results of the two simultaneous 2018 X-ray broadband spectra ({\sl XMM-Newton} and {\sl NuSTAR}) using this model: {\sc tbabs(reXcor + nthcomp + zgaussian)}.}
\centering
\begin{tabular}{@{}l c c}
\hline\hline
parameter & \multicolumn{1}{c}{2018 May}  & \multicolumn{1}{c}{2018 Nov.}\\
\hline
& \multicolumn{2}{c}{(a=0.99, h=5)}\\
\hline
$f_X$ ($\times$10$^{-2}$) &  3.9$\pm$0.1 & 3.0$^{+0.6}_{-0.2}$ \\
$\Gamma_{\rm hot}$ & 1.77$\pm$0.01   & 1.79$\pm$0.01 \\
$kT_{\rm hot} (keV)$ & 18$^{+4}_{-2}$ & 20$^{+20}_{-6}$  \\
$h_f$    & 0.45$^{+0.01}_{-0.02}$  & 0.51$^{+0.03}_{-0.02}$  \\
$\tau_{\rm warm}$  & 13.8$^{+0.3}_{-2.1}$  & 17.1$^{+1.0}_{-0.8}$  \\
log\,F(reXcor)$^{(a)}$ &  $-$11.26$\pm$0.01 & $-$11.26$^{+0.01}_{-0.03}$    \\
log\,F(nthcomp)$^{(a)}$ &  $-$10.73$\pm$0.01 & $-$10.65$\pm$0.01 \\
$\chi^{2}$/d.o.f.\ ($\chi^{2}_{\rm red}$) & \multicolumn{2}{c}{1871.5/1437 (1.30)}  \\
\hline
 & \multicolumn{2}{c}{(a=0.99, h=20)}\\
\hline
$f_X$ ($\times$10$^{-2}$)  &  $\leq$2.6 & $\leq$2.3 \\
$\Gamma_{\rm hot}$ & 1.76$\pm$0.01   & 1.79$\pm$0.01 \\
$kT_{\rm hot} (keV)$ & 16$^{+3}_{-2}$ & 18$^{+12}_{-4}$  \\
$h_f$    & 0.40$\pm$0.01  & 0.45$^{+0.01}_{-0.02}$  \\
$\tau_{\rm warm}$  & 11.5$^{+0.7}_{-0.4}$  & 15.0$\pm$0.8  \\
log\,F(reXcor)$^{(a)}$ &  $-$11.23$^{+0.01}_{-0.02}$  & $-$11.23$\pm$0.01    \\
log\,F(nthcomp)$^{(a)}$ &  $-$10.74$\pm$0.01 & $-$10.66$\pm$0.01 \\
$\chi^{2}$/d.o.f.\ ($\chi^{2}_{\rm red}$) & \multicolumn{2}{c}{1864.8/1437 (1.30)}  \\
\hline
& \multicolumn{2}{c}{(a=0.90, h=5)}\\
\hline
$f_X$  ($\times$10$^{-2}$)   &  2.8$^{+0.4}_{-0.5}$ & 7.4$^{+1.2}_{-0.9}$ \\
$\Gamma_{\rm hot}$  & 1.78$\pm$0.01   & 1.81$\pm$0.01 \\
$kT_{\rm hot} (keV)$ & 20$^{+6}_{-3}$ & 24$^{+68}_{-7}$  \\
$h_f$  & 0.57$\pm$0.02 & 0.67$\pm$0.01  \\
$\tau_{\rm warm}$  & 17.6$^{+0.1}_{-0.9}$  & 29.3$^{+0.7(p)}_{-1.0}$  \\
log\,F(reXcor)$^{(a)}$ &  $-$11.29$\pm$0.01 & $-$11.27$\pm$0.01    \\
log\,F(nthcomp)$^{(a)}$ &  $-$10.72$\pm$0.01 & $-$10.65$\pm$0.01 \\
$\chi^{2}$/d.o.f.\ ($\chi^{2}_{\rm red}$) & \multicolumn{2}{c}{1957.1/1437 (1.36)}  \\
\hline
&  \multicolumn{2}{c}{(a=0.90, h=20)}\\
\hline
$f_X$  ($\times$10$^{-2}$)   &  $\leq$3.8 & 4.2$^{+1.6}_{-1.2}$ \\
$\Gamma_{\rm hot}$ & 1.80$\pm$0.01  & 1.82$\pm$0.01 \\
$kT_{\rm hot} (keV)$ & 17$^{+4}_{-3}$ & 18$^{+16}_{-5}$ \\
$h_f$    & 0.58$^{+0.05}_{-0.04}$  & 0.68$^{+0.04}_{-0.02}$  \\
$\tau_{\rm warm}$  & 19.0$^{+1.9}_{-2.4}$  & 27.7$^{+1.2}_{-1.4}$  \\
log\,F(reXcor)$^{(a)}$ & $-$11.30$\pm$0.01 & $-$11.28$^{+0.01}_{-0.02}$   \\
log\,F(nthcomp)$^{(a)}$ &  $-$10.71$\pm$0.01 & $-$10.65$\pm$0.01 \\
$\chi^{2}$/d.o.f. ($\chi^{2}_{\rm red}$) & \multicolumn{2}{c}{1970.8/1437 (1.37)}  \\
\hline    \hline
\end{tabular}
\label{tab:rexcor}
\end{table}

\end{document}